\documentclass[12pt,preprint]{aastex}
\usepackage[dvips]{epsfig}
\usepackage[dvips]{color}

\begin{document}
\shorttitle{\sc ABUNDANCE SPREADS IN BO\"{O}TES~I AND SEGUE~1}
\shortauthors{NORRIS ET Al.}

\newcommand{\wcen} {$\omega$~Cen}
\newcommand{\boo}  {Bo\"{o}tes~I}
\newcommand{\boos} {Boo$-$1137}
\newcommand{\seg}  {Segue~1}
\newcommand{\segs} {Seg~1}
\newcommand{\uf}   {ultra-faint} 
\newcommand{\kms}  {km~s$^{-1}$} 
\newcommand{\lsun} {L$_{\odot,\rm{V}}$}
\newcommand{\msun} {M$_{\odot}$}
\newcommand{\teff} {$T_{\rm eff}$} 
\newcommand{\logg} {log~$g$} 
\newcommand{\g}    {$g$}
\newcommand{\gr}   {$(g-r)$}
\newcommand{\grz}  {$(g-r)_{0}$}
\newcommand{\rzz}  {$(r-z)_{0}$}
\newcommand{\bvz}  {$(B-V)_{0}$}
\newcommand{\rh}   {r$_{h}$} 

\title{CHEMICAL ENRICHMENT IN THE FAINTEST GALAXIES: THE CARBON AND
IRON ABUNDANCE SPREADS IN THE BO\"{O}TES~I DWARF SPHEROIDAL GALAXY AND
THE SEGUE~1 SYSTEM}

\author {JOHN E. NORRIS\altaffilmark{1}, ROSEMARY
F.G. WYSE\altaffilmark{2,6}, GERARD GILMORE\altaffilmark{3}, DAVID
YONG\altaffilmark{1}, ANNA FREBEL\altaffilmark{4}, MARK
I. WILKINSON\altaffilmark{5}, V. BELOKUROV\altaffilmark{3}, AND DANIEL
B. ZUCKER\altaffilmark{3,7,8}}

\altaffiltext{1}{Research School of Astronomy \& Astrophysics, The Australian National University, Mount Stromlo Observatory, Cotter Road, Weston, ACT 2611, Australia; email: jen@mso.anu.edu.au}
\altaffiltext{2}{The Johns Hopkins University, Department of Physics \& Astronomy, 3900 N.~Charles Street,  Baltimore, MD 21218, USA}
\altaffiltext{3}{Institute of Astronomy, University of Cambridge, Madingley Road, Cambridge CB3 0HA, UK}
\altaffiltext{4}{Harvard-Smithsonian Center for Astrophysics, Cambridge, MA 02138, USA}
\altaffiltext{5}{Department of Physics and Astronomy, University of Leicester, University Road, Leicester, LE1 7RH, UK.} 
\altaffiltext{6}{Institute for Astronomy, University of Edinburgh, Royal Observatory, Blackford Hill, Edinburgh EH9 3HJ, UK}
\altaffiltext{7}{Department of Physics, Macquarie University, North Ryde, NSW 2109, Australia}
\altaffiltext{8}{Anglo-Australian Observatory, PO Box 296, Epping, NSW 1710, Australia}
\begin{abstract}

We present an AAOmega spectroscopic study of red giant stars in
{\boo}, which is an {\uf} dwarf galaxy, and {\seg}, suggested to be
either an extremely low-luminosity dwarf galaxy or a star cluster. Our
focus is quantifying the mean abundance and abundance dispersion in
iron and carbon, and searching for distant radial-velocity members, in
these systems.

The primary conclusion of our investigation is that the spread of
carbon abundance in both {\boo} and {\seg} is large.  For {\boo}, four
of our 16 velocity members have [C/H] $\la$ --3.1, while two have
[C/H] $\ga$ --2.3, suggesting a range of $\Delta$[C/H] $\sim$
0.8. For {\seg} there exists a range $\Delta$[C/H]~$\sim$~1.0,
including our discovery of a star with [Fe/H] = --3.5 and [C/Fe] =
+2.3, which is a radial velocity member at a distance of 4 half-light
radii from the system center. The accompanying ranges in iron
abundance are $\Delta$[Fe/H]~$\sim$~1.6 for both {\boo} and {\seg}.
For [Fe/H] $<$ --3.0, the Galaxy's dwarf galaxy satellites exhibit a
dependence of [C/Fe] on [Fe/H] which is very similar to that observed
in its halo populations.  We find [C/Fe]~$\sim$ 0.3 for stars in the
dwarf systems that we believe are the counterpart of the Spite et
al. (2005) ``unmixed'' giants of the Galactic halo and for which they
report [C/Fe]~$\sim$~0.2, and which presumably represents the natal
relative abundance of carbon for material with [Fe/H] = --3.0 to
--4.0.
 
Our second conclusion is confirmation of the correlation between
(decreasing) luminosity and both (decreasing) mean metallicity and
(increasing) abundance dispersion in the Galaxy's dwarf
satellites. This correlation extends to at least as faint as M$_V$ =
--5, and may continue to even lower luminosities. The very low mean
metallicity of {\seg}, and the high carbon dispersion in {\boo},
consistent with inhomogeneous chemical evolution in near
zero-abundance gas, suggest these ultra-faint systems could be
surviving examples of the very first bound systems.

\end{abstract}


\section{INTRODUCTION}

The dwarf spheroidal (dSph) galaxies associated with the Milky Way
provide important potential insight into the $\Lambda$CDM paradigm,
the manner in which the baryonic material in low luminosity systems is
chemically enriched, and the formation of the halo populations of the
Galaxy (see Klypin et al.\ 1999; Moore et al.\ 1999; Tolstoy, Hill, \&
Tosi 2009, and references therein).  Recent studies of the newly
discovered {\uf}, high mass-to-light ratio, dwarf systems (e.g.\
Belokurov et al.\ 2006, 2007) place intriguing constraints not only on
the minimal baryonic masses with which a galaxy can form, but also on
the dark matter they contain and the interplay between dark and
luminous material in the production of the chemical elements at the
earliest times.  A crucial aspect of the ongoing discussion is the
nature of the lowest luminosity {\uf} systems.  For example, while
Geha et al.\ (2009) identify {\seg} as an {\uf} dwarf galaxy,
Niederste-Ostholt et al.\ (2009) have suggested instead ``it is a star
cluster, originally from the Sagittarius galaxy''.

The chemistry of the {\uf} systems is providing critical constraints
on their masses and their evolutionary histories, particularly by
focusing on the most metal-poor stars.  Kirby et al.\ (2008) first
reported stars with [Fe/H]\footnote{[Fe/H] =
log(N(Fe)/N(H))$_{\rm{star}}$ -- log(N(Fe)/N(H)$_{\odot}$)} as low as
--3.3, together with large abundance spreads, in several {\uf} dwarfs.
Norris et al.\ (2008), in a study of {\boo}, found a similar result,
with their most metal-poor star having [Fe/H] = --3.4.  Kirby et
al. (2008) also established that the mean metallicity of the dSphs
continues to decrease with declining luminosity, down to the faint
limit of the ultra-faint systems.

Frebel et al.\ (2010a) obtained the first relative abundances of
extremely metal-poor stars in the {\uf} systems, reporting not only a
range in Fe within each galaxy, but also in carbon, where two of three
stars in UMa II were found to have [C/Fe] = 0.5 and 0.8 at [Fe/H] =
--3.2.  Their work has been followed by further high-resolution,
moderate $S/N$ analyses of additional giants with [Fe/H] $<$ --3.0 in
the dwarf galaxies ({\seg}: Geha et al.\ 2009; Draco: Cohen \& Huang
2009; {\boo}: Feltzing et al.\ 2009, Norris et al.\ 2010; Sculptor:
Frebel, Kirby, \& Simon 2010b).  Perhaps the most interesting result
of these recent studies is that at [Fe/H] $\sim$ --3.7, the relative
abundances of a large number of elements are quite similar to those
found in the majority of Galactic halo giants (Frebel, Kirby, \& Simon
2010b (8 elements); Norris et al. 2010 (15 elements)).  A second
result is the report of surprisingly large values for the ratio of
abundances of two $\alpha$-elements, specifically [Mg/Ca], for one
star in Draco (Fulbright et al.\ 2004), two stars in Hercules (Koch et
al.\ 2008), and one star in {\boo} (Feltzing et al.\ (2009), most
simply interpreted in terms of inhomogeneous mixing of supernova
ejecta.

The present paper reports further results on the abundance ranges in
the {\uf} dwarf {\boo} (M$_{V,\,total}$ $\sim$~--6.3; Belokurov et
al.\ 2006, Martin, de Jong, \& Rix 2008) -- in particular, evidence
for a large range in the abundance of carbon -- together with evidence
for abundance spreads of both carbon and iron\footnote{Based on
analysis of the Ca II K line, and the assumption that [Ca/Fe] follows
the basic Galactic relationship.  See Section~5.} in {\seg}
(M$_{V,\,total}$ $\sim$~--1.5; Belokurov et al.\ 2007, Martin et al.\
2008).  Section~2 presents observational material, while Sections~3
and 4 present radial velocities of stars in the fields of {\boo} and
{\seg} and address the question of galaxy membership.  In Section~5,
we present atmospheric parameters {\teff}, {\logg}, and [Fe/H], while
in Sections 6 and 7 we reconsider the question of membership of two
apparently C-rich, extremely metal-poor stars in {\seg}, and present
relative carbon abundances, [C/Fe], for 16 radial-velocity candidate
members in {\boo} and three radial-velocity candidate members of
{\seg}.  The available data show that of four radial-velocity
candidate members of {\seg} for which data of sufficient quality are
available, one is carbon-rich and extremely metal-poor ([Fe/H] =
--3.5, [C/Fe] = +2.3), similar to the extremely rare carbon enriched
metal-poor (CEMP) stars having [Fe/H] $\la$ --3.5 in the Galactic
halo.  In Section~8 we discuss our results for abundance spreads and
dispersions and their implications for the formation, chemical
enrichment, and evolution of {\uf} galaxies. We show how the
comparably massive globular cluster $\omega$~Cen is consistent with
the inferred self-enrichment.  We continue the discussion and
summarize our results in Section~9.

\section{OBSERVATIONAL MATERIAL}

Candidate red giant members of {\boo} and {\seg} were observed with
the Anglo-Australian Telescope/AAOmega fiber-fed
spectrograph\footnote{See http://www.aao.gov.au/local/ www/aaomega/}
combination during 2007 April 18--20 and 2006 May 23--29 ({\boo} only;
the 2006 run was the first major visitor use of the new AAOmega
facility and these data sets were used to optimize and enhance the
data reduction system, {\tt 2dfdr}; final data calibration and
reduction used what is now the public {\tt 2dfdr} system).  This
instrument provides simultaneous spectra of 400 targets (science
targets plus dedicated sky fibers) over a field of 2~degrees in
diameter.  The light is split by a dichroic into blue and red regions
and sent to two separate spectrographs; only the blue spectra from the
2007 dataset will be discussed in this paper. These spectra have
resolution ${\cal{R}} = 5000$ and cover the wavelength range
3850--4540\ {\AA}.

Stellar targets for observation were selected from the SDSS DR4 data
set, based on their position in the relevant color-magnitude diagram,
using the selection masks from the discovery papers ({\boo}: Belokurov
et al.~2006; {\seg}: Belokurov et al.~2007).

\subsection{{\boo}}

\begin{figure}[htbp]
\vspace{1cm}
\begin{center}
\includegraphics[width=7cm,angle=00]{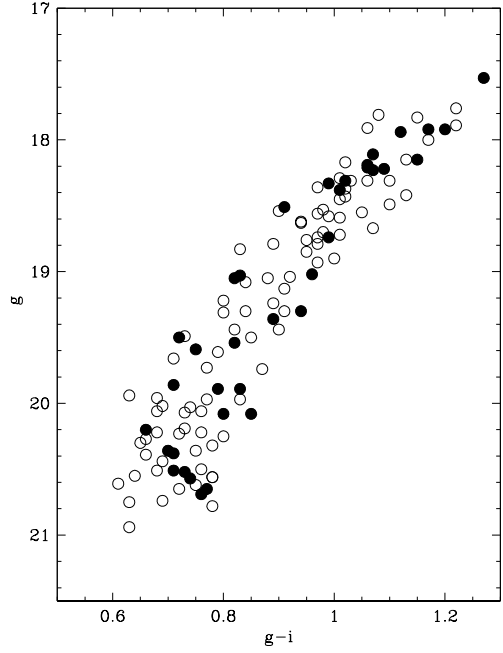}

\caption{\label{Fig:boo_cmd} Color-magnitude diagram of the target RGB candidates in {\boo} for which we obtained velocities. The filled
circles indicate radial-velocity candidate members.}

\end{center}
\end{figure}

\begin{figure}[htbp]
\vspace{1cm}
\begin{center}
\includegraphics[width=7cm,angle=270]{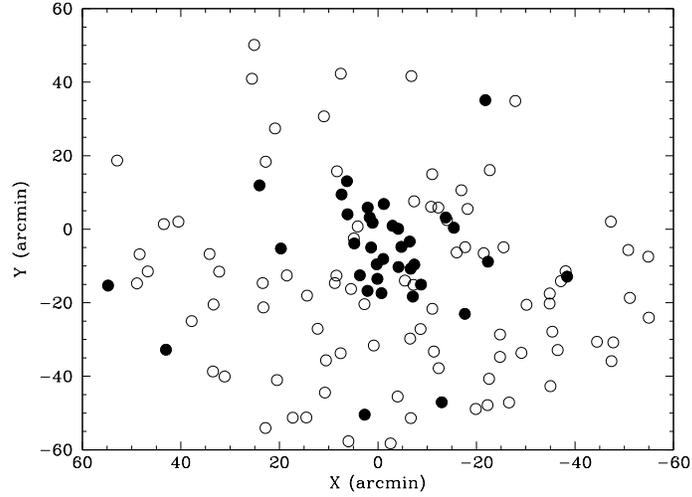}

\caption{\label{Fig:boo_XY} Distribution of RGB targets with
velocities, with respect to the nominal center of {\boo}, with offsets
in X and Y given in arcmin.  The symbol key follows that of
Figure~\ref{Fig:boo_cmd}, with filled symbols being candidate
radial-velocity members. Note that the nominal half-light radius is
13~arcmin (Belokurov et al.~2006).}

\end{center}
\end{figure}

We obtained useful spectra for stars in the magnitude range 17.5 $\la$
g $\la$ 21 (--1.6 $\la$ M$_{g}$ $\la$ 1.9 for a distance of 65~kpc
(Martin et al.\ 2008)).  The input target list included stars up to
one degree from the galaxy center, equivalent to four half-light radii
({\rh}) (Belokurov et al.~2006; Martin et al.~2008), and contained
blue-horizontal branch candidates (one of which proved to be a
quasar).  Only the stars lying on the red giant locus will be
discussed and analyzed in this paper.  The observed color-magnitude
diagram of such stars with velocity information is shown in
Figure~\ref{Fig:boo_cmd}, while the distribution on the sky is shown
in Figure~\ref{Fig:boo_XY}.

\subsection{{\seg}}

\begin{figure}[htbp]
\vspace{1cm}
\begin{center}
\includegraphics[width=7cm,angle=270]{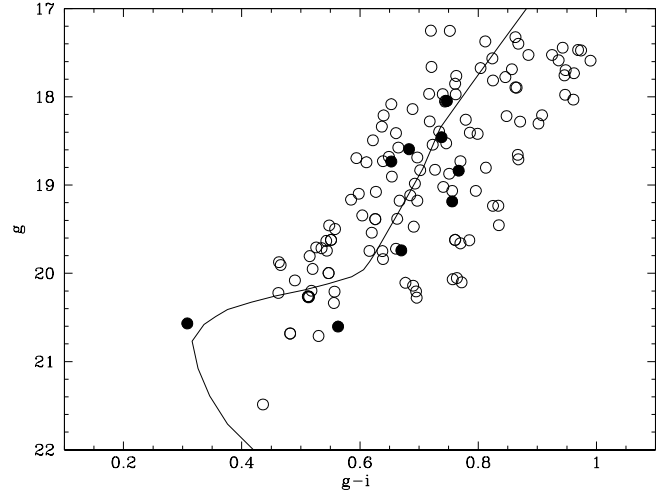}

\caption{\label{Fig:segue_cmd} Color-magnitude diagram of the target
stars in the RGB locus in {\seg} for which we obtained
velocities.  The filled symbols indicate candidate
radial-velocity members.  The fiducial of M92 (An et al.~2008),
adjusted to the appropriate distance modulus and reddening, is shown
as the smooth curve.}

\end{center}
\end{figure}

\begin{figure}[htbp]
\vspace{1cm}
\begin{center}
\includegraphics[width=7cm,angle=270]{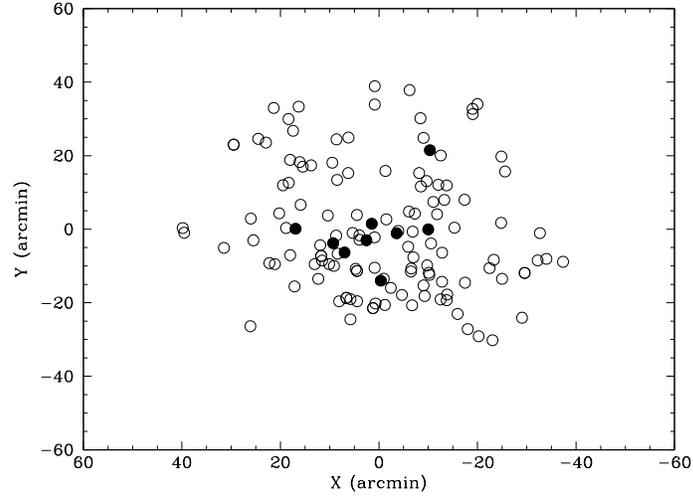}

\caption{\label{Fig:segue_XY} Distribution of targets in the RGB locus
for which we obtained velocities, with respect to the nominal center
of \seg, with offsets in X and Y given in arcmin.  Filled symbols
represent candidate radial-velocity members. Note that the nominal
half-light radius is 4.4~{\arcmin}.}

\end{center}
\end{figure}

We obtained useful spectra for stars in the magnitude range 17 $\la$ g
$\la$ 21.5 (0.4 $\la$ M$_{g}$ $\la$ 4.7 for a distance of 23~kpc
(Martin et al.\ 2008)).  We again selected targets many times the
nominal half-light radius (4.4~{\arcmin}, Martin et al.~2008) from the
center of this system, exploiting the wide-field capability of the
AAOmega system. Our target sample of 323 stars extended to 40{\arcmin}
from the galaxy center, corresponding to $\sim$~9{\rh}.  As for the
{\boo} field, the observed color-magnitude diagram of those stars for
which we obtained velocity information is shown in
Figure~\ref{Fig:segue_cmd}, while the distribution on the sky is shown
in Figure~\ref{Fig:segue_XY}.  Note the wide color range, due to the
poorly known location of the red giant branch of this system. {\seg}
is close enough that a direct comparison may be made with the fiducial
locus of the metal-poor globular cluster M92, derived from the SDSS
imaging data (An et al.~2008) and adjusted\footnote{We adopted the
distance modulus of 14.75 for M92, taken from Kraft \& Ivans (2003),
E(B$-$V) of 0.02~mag, and the extinction and reddening in the SDSS
filters calculated following An et al.\ (2008).} to the same reddening
and distance of {\seg}; this is the smooth curve in
Figure~\ref{Fig:segue_cmd}.

\section{RADIAL VELOCITIES AND MEMBERSHIP} 

Heliocentric radial velocities were determined using the {\tt HCROSS}
routine within the {\tt FIGARO} package (see
http://www.starlink.rl.ac.uk/star/docs/sun86.htx/node425.html). This
performs a cross-correlation between the program stellar spectrum and
a template, and determines the relative radial velocity.  An
associated confidence level and formal error are estimated (see
Heavens 1993); we accept only velocities with {\tt confidence = 1 or
1.0000} since experience shows those to have the cleanest
cross-correlation function and negligible rate of spurious results. We
excised the strong Ca II H \& K lines from the correlation analysis,
although this made an insignificant difference in most cases. Further,
there were defects on the CCD that affected a small wavelength range
in a subset of spectra and we also calculated cross-correlations with
that wavelength region (typically around 4380~\AA) excised.  The
G-giant standard star HD171391 (heliocentric radial velocity of
+6.9~{\kms}), for which spectra were obtained during the same
observing run as the {\uf} system candidates, was used as the template
since it provided more reliable cross-correlation than the alternative
twilight-sky template for lower signal-to-noise spectra\footnote{We
earlier reported, in Norris et al.~(2008) heliocentric velocities for
16 high signal-to-noise candidate members, using the twilight sky as a
template. The velocities reported here differ in the mean by only
1.25~{\kms}, with a dispersion of 3~{\kms}.}.  We calculated a
weighted mean (using the formal errors from the cross-correlation
package as weights) of the velocities from the two different
wavelength ranges, when both existed.  We have also removed a handful
of stars which turned out to be velocity-variable (binaries?), or
variable stars, or to have photometry inconsistent with being a single
star.  This resulted in a sample of 122 stars in the {\boo} field with
reliable velocities, and 134 in the {\seg} field.  The observational
data and derived heliocentric velocities are given in Table~1 ({\boo})
and Table~2 ({\seg}). The open symbols in Figure~\ref{Fig:boo_cmd} and
in Figure~\ref{Fig:boo_XY} indicate all those stars with reliable
velocities. The data were taken over several nights, with a wide
variety of sky conditions, with the resultant total integration times
in 2007, for example, being 9.5 hr and 7.5 hr, for {\boo} and {\seg},
respectively. The limitations of field acquisition, fiber placement
and weather mean that apparent magnitude is not a perfect predictor of
signal-to-noise.

Our internal accuracy on one observation, from repeat observations of
6 stars in the {\boo} field from 2006 and 2007 is 10~{\kms} with a
mean offset of 6~{\kms} if stars with {\tt confidence = 1.0000} are
included, and 7~{\kms} with a mean offset of $-1.5$~{\kms} if only the
4~stars with {\tt confidence = 1} are included\footnote{The difference
in the cross-correlation analysis between a confidence value of 1 and
one of 1.0000 is rather subtle.  Our experience with spectra of a
range of signal-to-noise has shown that the resulting velocities are
such that a value of 1.0000 tends to give higher formal errors, while
maintaining the same best estimate of the velocity, when compared to a
value of 1.}.  As noted in Norris et al.~(2008), external errors on
our radial velocities may be estimated from comparison with Martin et
al.~(2007). Using only the velocities relative to the G-star template,
HD171391, the six stars from our 2007 dataset in common with Martin et
al.~have a mean offset of $-2.7$~{\kms} and a dispersion of 13~{\kms};
this is dominated by one star, Martin et al.'s ID = 58, for which they
report an unusually large velocity error of 7.2~{\kms}, compared to
less than 2~{\kms} for the remaining 5 stars in common.  Removing that
star from the comparison gives a mean offset of 3~{\kms} and a
dispersion of 2.3~{\kms}.  This particular subsample has, on average,
higher signal-to-noise than is typical for our observations.  We
estimate from our repeats and standard-star observations that at
typical $S/N$ our velocities have combined internal and external
errors $\sim 10$~{\kms}.

\subsection{Membership}

\subsubsection{\boo}

\begin{figure}[htbp]
\begin{center}
\includegraphics[width=7cm,angle=270]{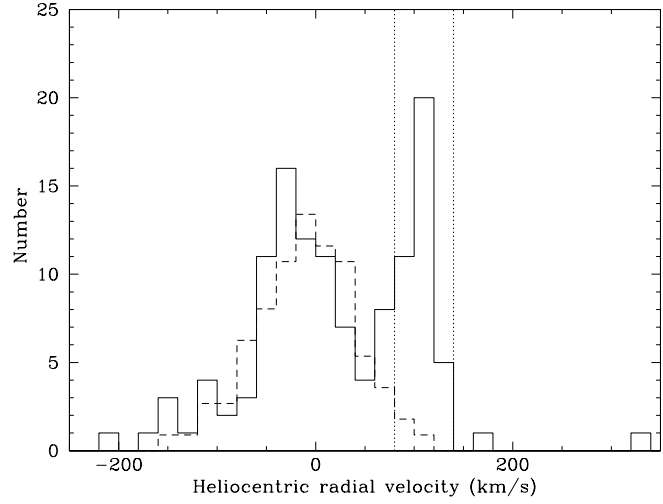}
\end{center}

\caption{\label{Fig:boo_vels} The solid histogram presents our derived
heliocentric radial velocities for the {\boo} field, while the dashed
histogram shows the predictions of the Besan{\c {c}}on model in this
line of sight, with our photometric selection and normalized to the
number of stars with reliable velocities in our sample that are not
candidate members of {\boo}\ (since this system is not included in the
Besan{\c {c}}on model). The local peak around $\sim 100$~{\kms} is due
to stars in {\boo}. The vertical dotted lines delineate our selection
range for candidate members.  Note that some field-star contamination
is expected.}

\end{figure}

Mu\~{n}oz et al.\ (2006) and Martin et al.\ (2007) reported a systemic
heliocentric radial velocity of $\sim 100$~{\kms}, with an internal
dispersion of $\sim 7$~{\kms}, for {\boo}.  The histogram of our
radial velocities is given in Figure~\ref{Fig:boo_vels}, with the
local maximum at around 100~{\kms} being due to members of {\boo}.
With random errors of the radial velocities of $\sim 10$~{\kms}, our
data are clearly incapable of resolving the internal kinematics of
{\boo}. (Our wide-field observations were designed to identify
candidate radial-velocity members out to large distances on the sky
for abundance study.)  {\boo}, at 65~kpc, is sufficiently distant
that interloper giant stars from the smooth stellar halo are
unexpected, but field main sequence stars could contaminate our
candidate members of {\boo}.  The line-of-sight is towards Galactic
latitude $\sim +70^\circ$, so that all Galactic components will have a
mean velocity close to zero. Star-count models, including the
Besan\c{c}on model of the Galaxy (Robin et al.~2003), predict that
stars in the Milky Way will have a distribution of heliocentric radial
velocities that peaks at $\sim -5$~{\kms}; matching our selection
criteria (as well as we can) to the Besan\c{c}on model interface
predicts that of Galactic stars, $\sim 6\%$ will be observed to have
heliocentric radial velocities at values higher than 75~{\kms}.  Thus
with our sample size of 122 stars, we expect of order eight Galactic
stars at high positive velocities.  The Besan\c{c}on model predictions
are indicated in Figure~\ref{Fig:boo_vels} and provide a satisfactory
match to the distribution at lower velocities.  The extra peak due to
the presence of {\boo} is quite pronounced; the thin vertical dotted
lines indicate the bin edges that contain our velocity range for
candidate radial-velocity members, $85 \leq V_{helio}$~({\kms}) $<
130$ (the bin edges are 80~{\kms} and 140~{\kms} and there are no
other stars with velocities in these bins). There are 36 stars in this
range, with a mean velocity of $ 105$~{\kms} and a dispersion of
$20$~{\kms}. There is a total of 40 stars with velocities above
75~{\kms}, so one might expect that of order four of the those in the
range we have selected as candidate radial-velocity members of {\boo}
are in fact Galactic contaminants, based on the Besan\c{c}on model
predictions.

The filled symbols in Figures~\ref{Fig:boo_cmd}
and~\ref{Fig:boo_XY} indicate our candidate radial-velocity members.
These candidates are also flagged in the final column in Table~1.

\begin{figure}[htbp]
\begin{center}
\includegraphics[width=7cm,angle=270]{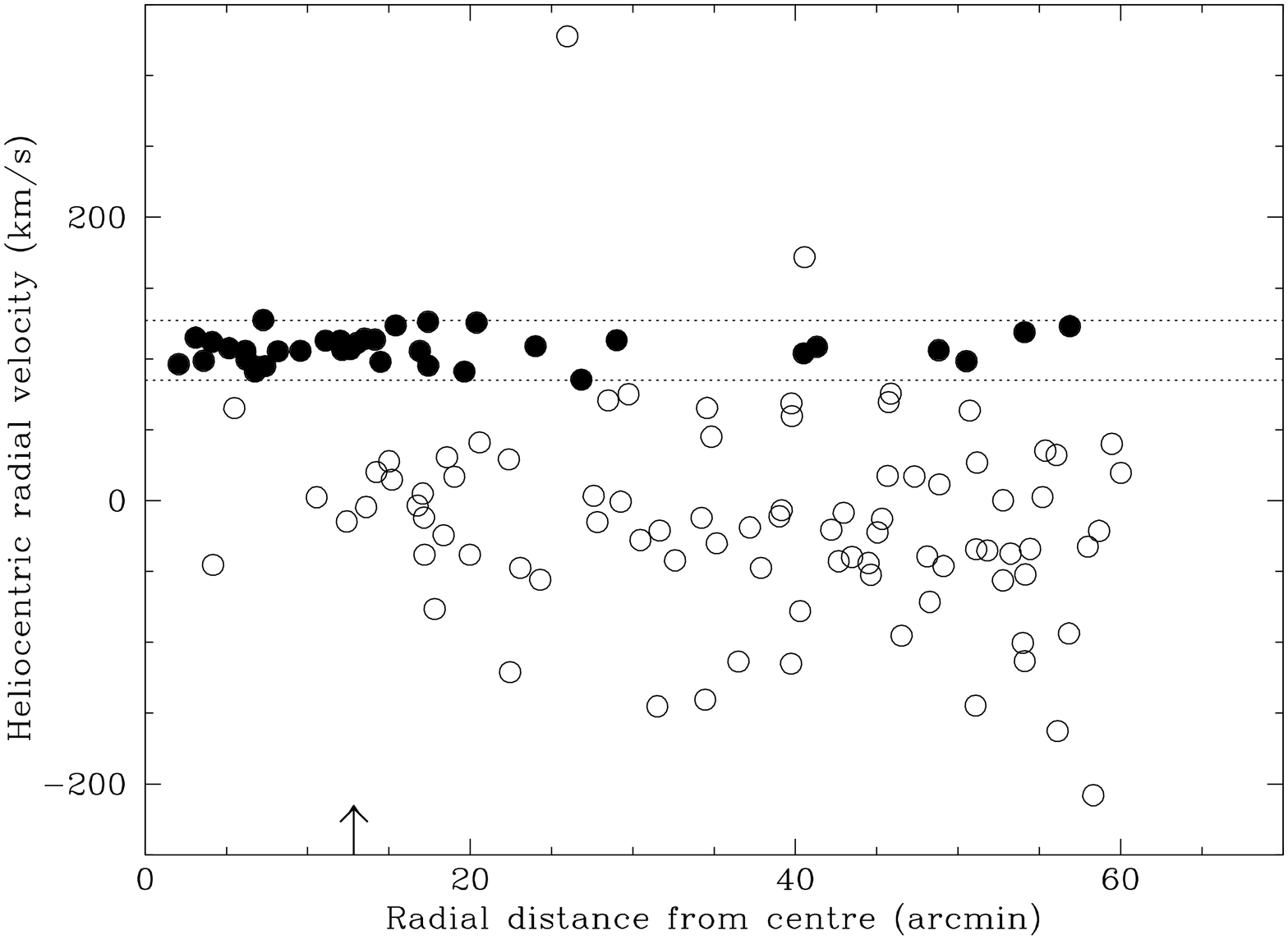}
\end{center}

\caption{\label{Fig:boo_vel_rad} Heliocentric radial velocity versus
projected radial distance from the center of {\boo}.  The velocity
range used to identify candidate members of {\boo}\ is indicated by
the dotted horizontal lines, while the filled symbols indicate the
candidate members. The half-light radius is indicated by the vertical
arrow on the x-axis.}

\end{figure}

Figure~\ref{Fig:boo_vel_rad} shows the distribution of velocities as a
function of projected radial distance from the galaxy's center.  It is
apparent from this figure that we identify radial-velocity members
well beyond the nominal half-light radius.  The outer parts of {\boo}
will likely be more susceptible to field contamination, and we can
investigate this by testing the observed distribution of velocities at
distant projected locations against a Gaussian, representing the field
Galaxy with no superposed dwarf galaxy. There are 56 stars that are
more distant than 35{\arcmin} (2.7 half-light radii, or 4.5
exponential scale-lengths) from the center of {\boo}, where there is a
gap in the distribution of candidate members in
Figure~\ref{Fig:boo_vel_rad}. These stars have a velocity distribution
that is in fact well-represented by a Gaussian, with mean velocity of
$-11$~{\kms} and dispersion of 75~{\kms} (not dissimilar to the
Besan\c{c}on predictions).  Adopting this Gaussian model, we calculate
the fraction of these stars which would lie by chance in the velocity
interval $85-130$~{\kms} (our selection for members of {\boo}) to be
7\%, or four stars out of our 56. The number of candidate
radial-velocity members of {\boo} in this range of parameter space is
six. That is, we have only moderate confidence in our detection of
member stars of {\boo} beyond 35{\arcmin}, based on velocity alone.
That said, high-resolution, relatively high-$S/N$ data recently
obtained for Boo--980, which lies at 3.9{\rh} and which we shall
present elsewhere (Frebel et al.\ 2010c, in prep.), lead to a (preliminary)
radial velocity of 99.0 $\pm$ 0.5 {\kms} (internal error) and
abundance [Fe/H] = --2.9 $\pm$ 0.2 that support its membership.

For completeness, repeating the fit to a Gaussian for the stars
projected within the inner 35~{\arcmin}, and fitting only to stars
outside the nominal velocity membership range, predicts three
contaminating field stars in the list of 30 candidate members. Again
this predicted total of seven contaminants agrees with the
Besan\c{c}on model predictions.

\subsubsection{\seg} 

\begin{figure}[htbp]
\begin{center}
\includegraphics[width=7cm,angle=270]{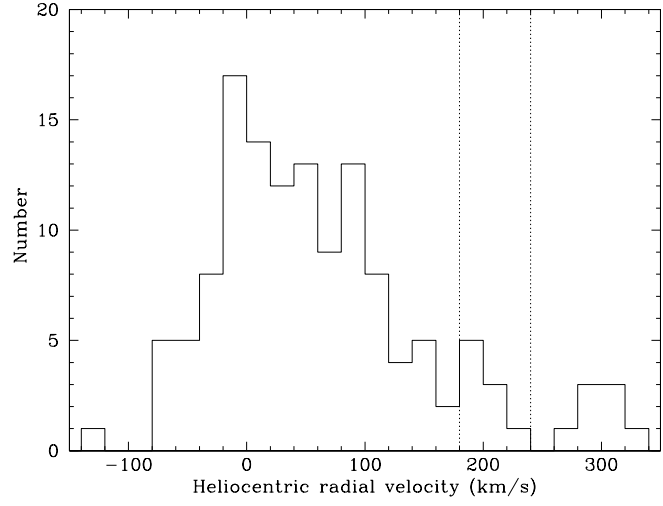}
\end{center}

\caption{\label{Fig:seg_vel_hist} The histogram shows our derived
heliocentric radial velocities for the {\seg} field, with the vertical
dotted lines, enclosing a local peak, delineating our selection range
for candidate members.  The local peak at $\sim 300$~{\kms} is of
unknown origin. }

\end{figure}

\begin{figure}[htbp]
\begin{center}
\includegraphics[width=7cm,angle=270]{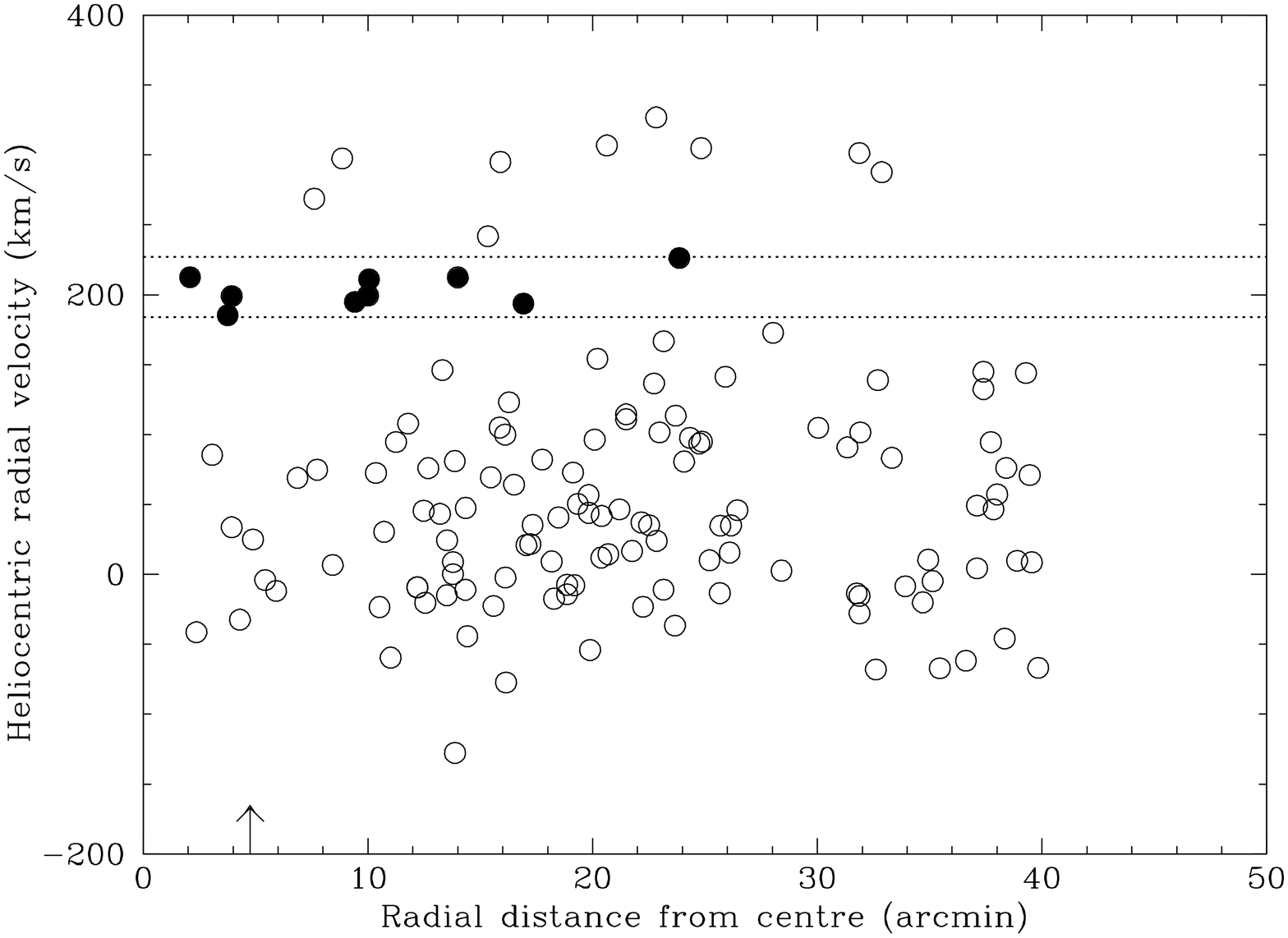}
\end{center}

\caption{\label{Fig:seg_vel_rad} Heliocentric radial velocity versus
projected radial distance from the center of \seg.  The velocity range
used to identify candidate members of \seg\ is indicated by the dotted
horizontal lines, and the filled symbols indicate the candidate
members. The half-light radius is indicated by the vertical arrow on
the x-axis.}

\end{figure}

We noted in Section 1 the competing suggestions of Geha et al.\ (2009)
and Niederste-Ostholt et al.\ (2009){\footnote { Niederste-Ostholt et
al.\ (2009) emphasize that {\seg} lies in a very complex part of the
outer Galaxy, not only projected onto the tidal stream of the
Sagittarius dSph, but also plausibly at the same distance
($\sim25$~kpc) with velocity similar to that of Sgr stream members.}
that the {\seg} system comprises either an ultra-faint dwarf or
material originating in the Sgr dSph.  Given the possibility that it
might actually comprise two distinct components, together with its
extremely small baryonic mass ($\sim$~1000{\msun}, Martin et
al. 2008), one might not be surprised to find that the establishment
of membership is problematic.  We shall see that this is indeed the
case.

The histogram of radial velocities for the {\seg} field is given in
Figure~\ref{Fig:seg_vel_hist}. Geha et al.\ (2009) reported a value of
206~{\kms} for the systemic radial velocity of {\seg}, with an
internal velocity dispersion of 4.3~{\kms}.  Again our data cannot
resolve the internal kinematics, and we may expect true members to be
scattered into $\pm 2\sigma$ of the systemic velocity, where here
$\sigma = 10$~km/s, our random error. Our velocity distribution shows
a reasonably well-defined local enhancement, of nine stars, between
185 $\leq v_{helio}$({\kms})~$<$ 230; these stars occupy the three
bins indicated in the histogram of Figure~\ref{Fig:seg_vel_hist}.  We
use this range to define our candidate members.  Extending the range
to 170 $< v_{helio}$({\kms})~$<$ 250 would add one star at each end,
for a total of 11 candidate radial-velocity members; the
plots here show only the nine candidates with 185 $\leq
v_{helio}$({\kms})~$<$ 230.  Figure~\ref{Fig:seg_vel_rad} shows the
distribution of velocities as a function of projected radial distance
from the system's center. It is apparent from these figures that we
identify candidate radial-velocity members well beyond the nominal
half-light radius.

The Galactic coordinates of {\seg} are ($\ell,\, b$) $ \sim $
(220$^\circ$, +50$^\circ$) and, as discussed in Geha et al.~(2009),
the velocity distribution of Galactic stars is expected to peak well
below the systemic velocity of {\seg}. The Besan{\c{c}}on
model predictions discussed in Geha et al.~(2009) lead to the
expectation that only 2.5\% of the total sample of Galactic stars
should have velocities in the range 190 $< v_{helio}$~({\kms})~$<$
220. However, the Besan{\c{c}}on model, which assumes smooth standard
kinematics for the field stars, is of limited usefulness in this
line-of-sight, given the known presence of the Sagittarius stream.

We may make a crude estimate of possible contamination directly from
our own velocity distribution function, since this presumably includes
some of the Sgr stream and other local halo structures which are not
included in the Besan\c{c}on model.  Our velocity distribution
declines roughly linearly in number, from the peak around 0~{\kms} to
close to zero objects at $\sim 200$~{\kms}.  Simply extrapolating that
decline under the velocity range of {\seg} is an uncertain process,
but suggests that up to four of the nine candidates could be field
contaminants.  Given the complexity of the local field, and the
possible similarity of kinematics between {\seg}, some part of the Sgr
streams, and possibly the nearby Orphan stream (Belokurov et
al.~2007), no robust {\it ab initio\/} spatial distribution model is
available to motivate a joint position-velocity membership criterion
(see Niederste-Ostholt et al.\ 2009, for an extended discussion).

Whatever {\seg} is or was associated with, it has the color-magnitude
diagram (CMD) of an old metal-poor population, and the derived
luminosity function above the turnoff and on the RGB must be
consistent with stellar evolution. Application of such consistency
checks is very uncertain.  Geha et al.~(2009) identify two (blue)
horizontal branch members and one might consider scaling from this to
the expected number of RGB members. However, as shown by
Niederste-Ostholt et al.\ (2009), there may be a significant number of
BHB interlopers from the Sgr stream, so the status of the two HB stars
ascribed by Geha et al.~to Segue~1 cannot be taken as assured.  The
total luminosity of {\seg} could in principle be used to predict the
number of members on the RGB, for example by assuming a stellar
population identical to that of the metal-poor globular cluster M3
(for which Renzini \& Fusi Pecci (1988) have tabulated the relative
memberships of different evolutionary stages). The spatial
distribution of our candidate members does not match well that used in
the estimate of the total luminosity (Belokurov et al.~2006), and
indeed the general membership uncertainties of stars in this line of
sight mean that the estimated luminosity itself is uncertain. Keeping
this complication in mind, Table 2 of Renzini \& Fusi Pecci shows that
M3 contains 342 RGB stars for every $ 3 \times 10^4$L$_\odot$, leading
to the expectation of 4 RGB stars for Segue 1, adopting a value of
350L$_\odot$. This estimate includes stars all the way down to the
base of the RGB, for which $M_g \sim +3.0$, or $g \sim 19.8$ for stars
at the distance of {\seg}.  Seven of our candidate radial-velocity
members are brighter than this limit, suggesting indeed that a
significant fraction, around half, are non-members. It is not possible
to say which stars are the contaminants and which are {\it bona
  fide\/} members. Although, as may be seen from Table~2, only 4/7 of
our candidates are within 3 half-light radii, using this information
involves assumptions about the (unknown) dynamical state of the
system. As we note below, our chemical abundance data do not allow
resolution of this uncertainty, but offer several possible interpretations
of the nature of {\seg}.

We complete this section by noting that there is also evident in
Figure~\ref{Fig:seg_vel_rad} a significant sample of stars with
velocity 300~{\kms}, which are distributed broadly across the field,
and which are not predicted by models with standard Galactic
kinematics.  Such a local peak in the velocity distribution was also
found by Geha at al.~(2009)\footnote{There is no overlap between our
sample and that of Geha et al.~(2009)}, who tentatively identify it
with tidal debris from the Sagittarius dSph. We discuss this stream
further using additional spectra in a separate paper (Frebel et al.\
2010d, in prep.).

\section{STARS WITH RELATIVELY HIGH-$S/N$ AAOMEGA SPECTRA}

Of some 98 {\boo} candidates having spectra with more than 200 counts per
0.34\ {\AA} pixel at 4150\ {\AA}, and which hence had sufficient $S/N$ for
a determination of metal abundance based on the Ca II K line strength
(see Norris et al.\ 2008), 16 fell in the radial-velocity range 90 $<$
V$_{r}$ $<$ 115~{\kms}, consistent with a high probability of being
{\boo} members.  Table~3 presents basic data for these objects, where
columns (1)--(3) contain the star name, radial distance from the
center of the galaxy, and radial velocity, respectively.  These stars
have a mean velocity of 105~{\kms}, and a dispersion of $6.5$~{\kms}.

We note here for completeness that had we increased the
radial-velocity limits for candidate membership to the range
85--130~{\kms} adopted above in Section 3.1.1, two further objects
would have been admitted as putative members.  These are Boo--2 and
Boo--71 in Table~1, which have velocities 123 and 91~{\kms}, and
distances from system center of 15.4{\arcmin} and 19.6{\arcmin}, respectively.
For Boo--71 we obtain [Fe/H] = --2.2, using the technique described in
Norris et al. (2008).  For Boo--2, however, our spectrum is of rather
poor quality and we hesitate to report an abundance -- very roughly we
estimate [Fe/H] $\sim$ --2.5.  While these objects are clearly worthy
of further consideration, we shall not discuss them further here.

For the {\seg} candidates, we shall consider here only those stars
having spectra with more than 200 counts per 0.34\ {\AA} pixel at
4150\ {\AA} and velocities in the range 185--230~{\kms} following the
discussion in Section~3.1.2.\footnote{For the larger velocity range
170--250~{\kms} considered in Section~3.1.2 one further putative
{\seg} member is admitted. {\segs}--117 in Table~2 has radial velocity
173~{\kms}, distance from galaxy center 28.0~{\arcmin}, and [Fe/H] =
--1.1.  We shall not consider this object further.} The data for the
five {\seg} candidates that meet these criteria are given in the first
five rows of Table~4 (which has the same column structure as Table~3).
Their spectra are presented in Figure~\ref{Fig:Segue1_Spectra} over
the wavelength range 3900--4400\ {\AA}.  One sees immediately that
these spectra are not what one might have expected for a sample of
objects taken from a stellar system with a monomodal abundance
distribution.  Two things are obvious.  First, there is a large range
in the strength of the G-band at 4300\ {\AA}; and second, the Ca II H
\& K lines (at 3968\ {\AA} and 3933\ {\AA}) differ widely from star to
star within the group.  In particular, {\segs}--7 and {\segs}--98 have
weak and somewhat ill-defined Ca II lines and very strong G-bands.
Such behavior is not unprecedented in extremely metal-poor stars of
the Galactic halo.  Figure~\ref{Fig:Segue1_UMP} compares the spectra
of these two stars with those of the halo, extremely metal-poor,
C-rich giants CS22949--037, CS29498--043, and HE0107--5240 (obtained
with the Double Beam Spectrograph on ANU's 2.3m telescope on Siding
Spring Mountain), which collectively have --5.4 $<$ [Fe/H] $<$ --3.8
and 0.9 $<$ [C/Fe] $<$ 3.8 (McWilliam et al.\ 1995; Aoki et al.\ 2002;
Christlieb et al.\ 2004). One consequence of the apparent
carbon-richness of the two {\seg} objects is that the contamination by
CH lines in the region of the Ca~K line may lead to erroneously
overestimated iron abundances derived from the K line on low
resolution spectra (see e.g. Christlieb et al.\ 2002; Frebel et al.\
2005; and Beers \& Christlieb 2005).

There are also two mundane explanations of the weak Ca~II lines in
{\segs}--7 and {\segs}--98 that one should consider.  The first is
that the effect is due to the relatively low $S/N$ of our spectra.
The second is that both stars are high-velocity (V$_{\rm r}$
$\sim$~200~{\kms}) stars with Ca II H \& K lines that have emission
cores\footnote{Given the decrease of Ca II H \& K lines emission with
increasing age (at least in dwarfs; see Barry 1988), the high
velocities (and presumably large ages) of the {\seg} objects might
suggest less contamination from this source.}. We searched our
collection of some 3000 medium resolution (FWHM $\sim$ 2.5\ {\AA})
spectra of Hamburg ESO Survey (HES) metal-poor candidates (see Norris
et al.\ 2007) for stars that have weak Ca II H \& K lines as the
result of core emission, and found some 10 objects with relatively
weak H \& K emission leading to weak overall H \& K line absorption.
In Figure~\ref{Fig:Segue1_HKem} we compare the spectra of {\segs}--7
and {\segs}--98 with four of these objects.  There is clearly not a
good match between the spectra of the HES stars and those of the
candidate {\seg} stars. However, one might imagine that if in the
candidate {\seg} stars there were weaker emission features than those
seen in the HES stars in Figure~\ref{Fig:Segue1_HKem} the abundances
we derived for the candidate {\seg} stars could be erroneous.

\section{ATMOSPHERIC PARAMETERS}

To determine relative carbon abundances ([C/Fe]) in what follows, we
shall also need the atmospheric parameters effective temperature
({\teff}), surface gravity ({\logg}), and metal abundance ([M/H]),
where for simplicity we shall assume [M/H] = [Fe/H].  In Tables 3 (for
{\boo}) and 4 (for {\seg}) we present data that we have used for this
purpose.  Columns (4)--(6) contain $ugriz$ photometry for $g$, {\grz},
and {\rzz} from Data Release 7 of the Sloan Digital Sky Survey
(Abazajian et al.\
2009\footnote{\texttt{http://cas.sdss.org/astrodr7/en/tools/search/}}),
where the colors have been corrected for reddenings corresponding to
E$(B-V)$ = 0.02 (Belokurov et al.\ 2006) and 0.032 (Geha et al.\
2009), for {\boo} and {\seg}, respectively.  From our spectra of the
{\seg} objects in Table~4, we measured values of the Ca II K
line-strength index, K$^{\prime}$, defined by Beers et al.\ (1999),
which are presented in column (7) of Table~4.  For {\boo},
K$^{\prime}$ values from Norris et al.\ (2008) are included in column
(7) of Table~3.

As described in Norris et al.\ (2010), {\teff} and {\logg} can be
estimated for metal-poor red giants in dSph systems by using $ugriz$
photometry together with the synthetic $ugriz$ colors of
Castelli\footnote{\texttt{http://wwwuser.oat.ts.astro.it/castelli/colors/sloan.html}}
and the Yale--Yonsei (YY) Isochrones (Demarque at al.\ 2004\footnote{
\texttt{http://www.astro.yale.edu/demarque/yyiso.html}}) with an age of
12 Gyr, and the assumption that the stars lie on the red giant branch
of the system.

For the {\boo} and {\seg} stars investigated here, values of {\teff}
and {\logg} were obtained for each of {\grz} and {\rzz}, and also
{\bvz} (see below).  A check on these atmospheric parameters was
obtained by using a technique similar to that described by Norris et
al.\, in which the absolute magnitude M$_{V}$ -- derived from the
apparent magnitude and distance modulus of the parent system --
replaces color, together with the adoption of the Yale-Yonsei
Isochrones and the assumption that the stars lie on the red giant
branch of the parent dSph.  To determine the estimated values of
absolute visual magnitude M$_{V}$, we used the Lupton\footnote
{\texttt{http://www.sdss.org/dr5/algorithms/sdssUBVRITransform.html\#Lupton2005}}
(V, {\g})-- transformation, together with our adopted reddening, the
distance moduli of Martin et al.\ (2008), and the assumption
A$_{V}$~=~3~$\times$~E($B-V$).  Then, for the YY isochrone of assumed
abundance [Fe/H] (and the above age of 12 Gyr), by interpolation in
the ({\teff}, M$_{V}$)-- and ({\logg}, M$_{V}$)-- relationships
defined by the isochrone we determine the values of {\teff} and
{\logg} corresponding to the derived value of M$_{V}$.  Agreement
between the atmospheric parameters obtained from $ugriz$ colors with
those from absolute magnitude for {\boo} was excellent: for the 16
stars in Table~3 we obtained average differences
$\langle$$\Delta${\teff}$\rangle$ = 70K and
$\langle$$\Delta${\logg$\rangle$} = 0.2.  For {\seg}, on the other
hand, the agreement was considerably poorer, with
$\langle$$\Delta${\teff}$\rangle$ = 230K and
$\langle$$\Delta${\logg}$\rangle$ = 0.5.  If, however, we exclude
{\segs}--42, the hottest star in the sample, and for which the two
{\teff} and {\logg} estimates differ by 440K and 0.9 dex respectively,
we obtain $\langle$$\Delta${\teff}$\rangle$ = 140K and
$\langle$$\Delta${\logg}$\rangle$ = 0.4.  We shall return to this
point below, once we have discussed the abundances of the putative
{\seg} members.

Metal abundances were determined in two ways.  First, we used the
calibration of Beers et al.\ (1999), which permits one to determine
[Fe/H], given observed values of K$^{\prime}$ and
(B--V)$_{0}$\footnote {The reader should be aware the Beers et al.\
(1999) calibration assumes that the same [Ca/Fe] vs [Fe/H]
relationship applies within both the metal-poor Galactic calibration
objects and the dSph satellites.  While this is not true at higher
abundances, it appears to hold for [Fe/H] $<$ --2.0 (e.g. Scl: Tolstoy
et al.\ 2009; UMa II and Com: Frebel et al.\ 2010a; {\boo}:
Norris et al.\ 2010). }.  In order to do this, we determined
(B--V)$_{0}$ from the values of {\grz} in Tables 3 and 4.  For stars
with {\grz} $>$ 0.545 (corresponding to $(B-V)_{0}$ $>$ 0.7), we
adopted the transformation {\bvz} = 1.197$\times${\grz} + 0.049,
appropriate for metal-poor red giants, following Norris et al.\
(2008)\footnote{Our quoted [Fe/H] values for {\boo} differ trivially
from those of Norris et al.\ (2008) because of the small differences
between SDSS DR4 and DR7.}.  This applied for all but two objects
({\segs}--31 and {\segs}--42), for which we adopted {\bvz} =
0.916$\times${\grz} + 0.187, from Zhao \& Newberg (2006), valid for
metal poor-stars over the range --0.15 $<$ {\grz} $<$
0.55.  

The second method of abundance determination involves model atmosphere
analysis of high-resolution, high $S/N$, spectra obtained with
VLT/UVES (Norris et al.\ 2010, for {\boos}), and with VLT/UVES/Flames
for a further seven of the {\boo} stars, which will be reported
elsewhere (Gilmore et al.\ 2010 in prep.).

The atmospheric parameters for {\boo} and {\seg} are presented in
Table~3 and 4, where columns (8)--(9) contain {\teff} and {\logg}
obtained from colors as described above, while column (10) presents
[Fe/H] (except for {\segs}--7 and {\segs}--98}), based on the above
two methods, where the tabulated abundance gives precedence to the
UVES based value if available, failing which the AAOmega
K$^{\prime}$-based result is used.
 
Our abundances for {\seg} deserve comment.  For {\segs}--31 and
{\segs}--71, we derive relatively high values of [Fe/H] = --1.9 and
--2.2, respectively.  To give the reader a feeling for the case for
the high abundances of these objects, we compare their spectra in
Figure~\ref{Fig:Segue1_normal} with those of Galactic halo giants
having similar colors, and abundances in the range --4.8 $<$ [Fe/H]
$<$ --2.2.  (The colors of the Galactic stars are taken from Cayrel et
al.\ (2004), Norris, Bessell, \& Pickles (1985), and Norris et
al. (2007), while their abundances come from Cayrel et al.\ (2004),
Chiba \& Yoshii (1998), and Norris et al.\ (2007).)  Examination of the
spectra, in the region of the Ca II H \& K lines (3900--4000\ {\AA}),
clearly supports our relatively high abundances.  That said, we recall
that Geha et al.\ (2009) have reported a red giant in {\seg} (their
star 3451364) with [Fe/H] = --3.3.  We note then that the {\seg}
spectra in Figure~\ref{Fig:Segue1_normal} have much stronger Ca II H
\& K lines than seen in CS22897--008 (in the third panel from the
top), which has [Fe/H] = --3.4, similar to the abundance of the Geha
et al. star.  Said differently, there appears to exist a large
abundance range ($\Delta$[Fe/H] $\sim$~1~dex) within the sample of
candidate members of {\seg}.

The abundances of the more-enriched candidate members of {\seg} are,
however, close to that of a typical field halo star at the distance of
{\seg} (the outer halo of Carollo et al.\ 2007). Further, there are
clear indications of an abundance gradient from the core of the Sgr
dSph to its tidal streams, with a mean metallicity [Fe/H]~$\sim$~--1
derived from M giants (at heliocentric distances greater than $\sim
10$~kpc; Chou et al.\ 2007) and --1.8 from RR Lyrae stars in the
leading arm, at heliocentric distances of $\sim 50$~kpc (Vivas, Zinn,
\& Gallart 2005). These results suggest that at [Fe/H] $\sim$ --1.5 --
--2, both field halo stars and Sgr debris would be hard to distinguish
from members of {\seg}, given that some models of the Sgr streams
predict similar velocities (e.g.~Niederest-Ostholt et al.\ 2009).  

For the two putative C-rich objects in {\seg}, we also derived [Fe/H]
using the Beers et al.\ (1999) formalism, which leads to ${\rm [Fe/H]
= -3.6}$ and --4.0 for {\segs}--7 and {\segs}--98, respectively.  At
this stage we regard these values as uncertain for the reasons
discussed in Section~4. We shall defer further discussion of them
until Section~6.

It is somewhat difficult to estimate realistic errors for {\teff} and
{\logg} that include both internal and potential external errors.  The
cited errors in the DR7 $ugriz$ colors propagate to relatively small
errors in {\teff} and {\logg}, and it is probably more realistic to
concentrate on systematic effects.  One measure of this could be the
internal spread in the results of the three methods based on
calibrations of {\grz}, {\rzz}, and {\bvz} described above.  We find
that the averages of the standard error of the mean for the 16 {\boo}
objects in Table~3 are $\langle$$\Delta${\teff}$\rangle$ = 40K and
$\langle$$\Delta${\logg}$\rangle$ = 0.1.  These are somewhat smaller
that the average differences of $\langle$$\Delta${\teff}$\rangle$ =
70K and $\langle$$\Delta${\logg}$\rangle$ = 0.2 obtained above from
colors and absolute magnitude (i.e.~assuming that the stars are indeed
red giants at the distance of {\seg}).  We adopt the larger
differences as error estimates in what follows.  For the five {\seg}
stars we find average differences of $\langle$$\Delta${\teff}$\rangle$
= 90K and $\langle$$\Delta${\logg}$\rangle$ = 0.2; and if we exclude
{\segs}--42, the problematic star discussed above, these values become
$\langle$$\Delta${\teff}$\rangle$ = 100K and
$\langle$$\Delta${\logg$\rangle$} = 0.3.

We now return to the issue of the discrepant values of {\teff} and
{\logg} obtained above for {\segs}--42 from estimates based on color
and absolute magnitude.  While this object appears to be the hottest of
the stars in Table~4, it is also the most metal-rich, with
[Fe/H] = --1.5.  Given that this is fundamentally at odds with basic
concepts of evolution on the red giant branch (more metal-rich stars
should be cooler), and our assumption that {\segs}--42 lies on the
giant branch of {\seg}, we shall exclude {\segs}--42 from further
consideration, on the assumption it is not a red giant member of
{\seg}.  In what follows we shall adopt mean errors of
$\sigma$({\teff}) = 140K and $\sigma$({\logg}) = 0.4 for the four
remaining candidate {\seg} giants, the larger of our two estimates
obtained when we exclude {\segs}--42 from consideration.  We note
parenthetically here that the differences between our two methods of
deriving atmospheric parameters are commensurate, to within these
errors, with the assumption that the other stars in Table~4 lie on the
RGB of {\seg}. 

Abundance errors for {\boo} are $\sigma$[Fe/H] = 0.35 for data
obtained with AAOmega (Norris et al.\ 2008) and $\sigma$[Fe/H] = 0.15
for the results from VLT/UVES and VLT/Flames/UVES (Norris et al. 2010;
Gilmore et al.\ 2010 in prep.)\footnote{We note for completeness that for
[Fe/H] the mean difference and the dispersion of differences for the
seven stars in our Table~3 having both AAOmega and VLT/Flames
abundances are --0.06 and 0.44~dex, respectively. For the four stars
in common between our high-resolution VLT/Flames abundances in
Table~3, and those from the Keck/HIRES spectra of Feltzing et al.\
(2009) the corresponding values are --0.01 and 0.13 dex.}.  For
{\segs}--31 and {\segs}--71 we adopt $\sigma$[Fe/H] = 0.4.

 \section {FURTHER ASSESSMENT OF THE PUTATIVE SEGUE~1 C-RICH, EXTREMELY METAL-POOR STARS}

It is difficult to over-emphasize the implications of {\seg}
membership of the two C-rich, extremely metal-poor candidates
{\segs}--7 and {\segs}--98.  Put most simply, C-rich stars, having
[Fe/H] $\la$ --3.3 are extremely rare: few are known, and only some
seven have been analyzed in detail\footnote{We refer to CS22949--037
(McWilliam et al.~1995), CS22957--027 (Norris et al.~1997),
CS29498--043 (Aoki et al.~2002), HE0107--5240 (Christlieb et al.\
2004), HE0557--4840 (Norris et al.~2007), HE1327--2326 (Frebel et
al.~2005), and G77--61 (Plez \& Cohen 2005).}. The existence of such
objects in the {\uf} dwarfs would have strong implications for these
systems being building blocks of the Galaxy's outer halo.

After the analysis reported here was complete, we sought to obtain
high-resolution, high $S/N$ data for both {\segs}--7 and {\segs}--98.
There were two significant developments.  First,  Evan  Kirby and
Joshua Simon informed us that they have near-infrared spectra for
{\segs}--98 that yield much higher values of [Fe/H] than suggested
here, and generously permitted us to examine several of their spectra.
We accept their argument, and shall not consider that object further
in this paper. Further data are needed to clarify the inconsistency.

Second, in April 2010, we obtained spectra taken with VLT/UVES of
{\segs}--7 with resolution R = 35,000, and $S/N$ = 40/0.027\ {\AA}
pixel at 4300\ {\AA}.  We shall report the analysis of these data at a
later time.  Suffice it here to say that our preliminary analysis
yields a radial velocity of 204.3 $\pm$ 0.1 {\kms} (internal error),
while for atmospheric parameters {\teff} = 4960K and {\logg} = 1.9
(from Table 4), and using techniques described in Norris et al.\
(2010), our preliminary analysis yields [Fe/H] = --3.5, and
[C/Fe]~$\sim$~+2.5 for this object.  We shall retain {\segs}--7 in our
analysis, and list this value of [Fe/H] (which agrees well with that
reported in the previous section) in Table 4.  Finally, for
completeness and the discussion that follows, we also include the data
of Geha et al.\ (2009) for their red giant 3451364 in the final row of
Table 4.

\section {CARBON ABUNDANCES}

AAOmega spectra of the {\boo} members in Table~3 were presented in the
wavelength range 3900--4400\ {\AA} in Figure 1 of Norris et al.\ (2008),
to which we refer the reader.  In that figure one sees a clear and
relatively large range in the strength of the G-band (of the CH
molecule) at 4300\ {\AA}.  We analyze these band strengths below to
determine carbon abundances.

Using the atmospheric parameters in Tables 3 and 4, we have computed
(1D/LTE) model atmosphere synthetic spectra in the region of the
G-band 4270--4335\ {\AA} using the procedures described by Stanford et
al.\ (2007, and references therein), to which we refer the reader for
details. Suffice it here to say that the spectrum synthesis code was
that of Cottrell \& Norris (1978), the models used were those of
Kurucz (1993), and the CH $gf$ values of Kurucz were modified by 0.5
dex to provide a fit between the observed and synthetic model
atmosphere spectrum of the Sun.  As a further test of the procedure,
we determined the relative carbon abundance for the archtypical
metal-poor red giant HD 122563 using the observed spectrum and
atmospheric parameters ({\teff} = 4650K, {\logg} = 1.4, [Fe/H] =
--2.7) of Ryan et al.\ (1996) and adopting {[O/Fe] = +0.6\footnote{In
the absence of information on the oxygen abundance in {\boo} and
{\seg}, we have also assumed [O/Fe] = +0.6 in the present
investigation, which is typical of metal-poor stars in the Galactic
halo, throughout this investigation.  Some support for this comes from
the fact that available $\alpha$-element abundances show consistency
between most metal-poor dwarf galaxy members and the Galactic halo.}.
The value obtained was [C/Fe] = --0.35, in good accord with the value
of --0.4 reported by Sneden (1973) and by Cayrel et al.\ (2004).

The fits of the synthetic spectra to the present observations are
shown in Figure~\ref{Fig:Bootes_Carbon} for {\boo}, for the 12 stars
in Figure 1 of Norris et al.\ (2008) (there presented over the larger
range 3900--4400\ {\AA}) and in Figure~\ref{Fig:Segue_Carbon} for
{\seg}. The derived abundances, [C/H], are presented in column (11) of
Tables 3 and 4.  Their accuracy is estimated to be ${\Delta}$[C/H] =
0.2 and 0.4}, for {\boo} and {\seg}, respectively, which include the
propagation of errors in the atmospheric parameters {\teff}, {\logg},
and [Fe/H] given in Section~5 above (of which the first is the most
significant), and the fitting error.  [C/H] for {\boos} in Table~3 has
also been determined by Norris et al.\ (2010) using high-resolution,
high $S/N$ VLT/UVES spectra, interpreted using the MOOG code of Sneden
1973) (see Norris et al.\ 2010 for details) and the models of Castelli
\& Kurucz (2003;
\texttt{http://wwwuser.oat.ts.astro.it/castelli/grids.html}).  They
determined [C/H]~$ = -3.41$, in good accord with the value of
[C/H]~=~--3.46 in Table~3.

\section{ABUNDANCE SPREADS AND DISTRIBUTIONS}

\subsection{Carbon Abundances in the Dwarf Spheroidal Galaxies of the Milky Way}
\subsubsection{The Spread of [C/H] in {\boo} and {\seg}}

The first conclusion of our investigation is that the spread of carbon
abundance in both {\boo} and {\seg} is large.  For the former, four of
the 16 objects in Table~3 have [C/H] $\la$ --3.1, while two have [C/H]
$\ga$ --2.3, suggesting a range of $\Delta$[C/H] $\sim$ 0.8. In
Table~4 one finds that for {\seg} there exists a range
$\Delta$[C/H]~$\sim$~1.0.  These results are relatively robust to
errors in the atmospheric parameters assumed in the analysis: for
example, representative errors of $\Delta${\teff} = 100K,
$\Delta${\logg} = 0.3, and $\Delta$[Fe/H] = 0.3, cause errors in [C/H]
of 0.2, 0.1, and 0.0, respectively.  (The ranges in [C/Fe] in {\boo}
and {\seg} $\Delta$[C/Fe] are $\sim$~1.0 and $\sim$~2.5, respectively.
We discuss the dispersion in [C/Fe] below.)

\subsubsection{Relative Carbon Abundance, [C/Fe], as a Function of Metallicity 
and Absolute Magnitude}
   
In Figure~\ref{Fig:[C/Fe]_[Fe/H]} we compare the behavior of relative
abundance [C/Fe] as a function of [Fe/H] for {\boo}, {\seg}, other
dwarf systems, and for the Galactic halo.  In the upper panel the open
and filled stars represent red giants in {\boo} and {\seg},
respectively, while the filled green circles stand for stars in the
{\uf} dwarfs UMa~II and Com from Frebel et al.\ (2010a) and the filled
blue ones for stars in the more luminous Draco dSph (from Cohen \&
Huang (2009).  The lower panel presents results for red giants in the
Galactic halo taken from Spite et al.\ (2005), where open and filled
red circles denote their ``mixed'' and ``unmixed'' stars. As we shall
see in Figure~\ref{Fig:[C/Fe]_Mv} the incidence of ``mixed'' stars
increases with increasing luminosity.  Filled black circles represent
stars with large enhancements of C and N (and often O) from Aoki et
al.\ (2002), Cayrel et al.\ (2004), Christlieb et al.\ (2004), Norris,
Ryan, \& Beers (1997), and Norris et al.\ (2007) (see the figure
caption for identifications). While there is a clear similarity
between the two panels, some caution is warranted.  The sample in the
lower panel of the figure is not unbiased.  Of the 39 objects plotted,
30 come from the HK survey (Beers, Preston, \& Shectman 1992), two
from the HES (HE0107--5240 (Christlieb et al.\ 2004) and HE0557--4840
(Norris et al.\ 2007), both with [Fe/H] $<$ --4.4), and five from
brighter miscellaneous sources (all with [Fe/H $>$ --3.0).  There have
been strenuous endeavors to observe all stars in the HK survey with
[Fe/H] $<$ --3.0: to our knowledge, as a result of this comprehensive
effort, some 35 of the red giants in the HK survey in the range --4.2
$<$ [Fe/H] $<$ --3.0 now have high quality abundance analyses, 23 of
which appear in Figure~\ref{Fig:[C/Fe]_[Fe/H]}.  Given the fact that
Spite et al. (2005) chose to observe stars with [Fe/H] $<$ --2.5, the
data in the lower panel are therefore very incomplete above that
limit.  In contrast, the two HES stars in the figure, with [Fe/H] $<$
--4.4, represent only the (extremely important) low abundance tail of
the HES distribution.

Figure~\ref{Fig:[C/Fe]_Mv} presents the behavior of [C/Fe] as a
function of absolute visual magnitude, M$_{V}$.  The upper panel (a)
presents results for the red giants in {\boo}, {\seg}, and dwarf
galaxies as defined in Figure~\ref{Fig:[C/Fe]_[Fe/H]}, while panels
(b)--(d) contain data for three other stellar populations of the Milky
Way.  Panel (b) presents giants of the Galactic halo (the sample
presented in Figure~\ref{Fig:[C/Fe]_[Fe/H]}.  Note that the incidence
of ``mixed'' stars of Spite et al.\ (2005) (the open symbols)
increases as one moves to higher luminosity, consistent with the
interpretation of increased importance of mixing as the star ascends
the giant branch, crossing the ``red giant branch bump'' (see Gratton
et al.\ 2000).  Panel (c) contains giants in the globular clusters M15
(open circles) and M92 (filled circles) (Langer et al.\ 1986; Trefzger
et al.\ 1983), both of which have [Fe/H] $\sim$ --2.2, with little or
no internal dispersion in iron abundance.  Finally, (d) presents
results for the brightest giants in the Galaxy's most massive and
chemically inhomogeneous globular cluster $\omega$~Centauri (from the
chemically biased sample of Norris \& Da Costa 1995, hereafter
ND). (Open and filled circles refer to CO-weak and CO-strong objects
(of which more below), while the asterisk represents a CH star.)
{\wcen} has been proposed to have once been the core of a nucleated
dwarf galaxy, captured long ago by the Milky Way (see Bekki and
Freeman 2003, and references therein) so that the spread of chemical
abundances reflects evolution within the deeper potential well of the
host galaxy (e.g. Bekki and Norris 2006; Marcolini et al.\ 2007).  If,
alternatively, {\wcen} itself possessed a high enough baryonic mass to
retain some ejecta from evolved stars and supernovae (e.g.~Norris et
al. 1996; Gnedin et al.\ 2002) it is simply another example of a
self-enriching system and would be expected to show similar scalings
to those of the dwarf galaxies. We favor this second model below.

One sees a large range of [C/Fe] in these panels.  For heuristic
purposes we also show the continuous line corresponding to the
division of Aoki et al.\ (2007) between C-rich and C-normal stars of
the Galactic halo, taking account of mixing, and associated carbon
reduction, as stars ascend the RGB.

\subsubsection{Red Giants as Probes of Galactic Carbon Enrichment}

The data in Figure~\ref{Fig:[C/Fe]_Mv} encapsulate the difficulties
one encounters in seeking to interpret the observed carbon abundances
of red giants in the old stellar populations of the Milky Way system
in terms of galactic chemical enrichment.  Not only must one
understand the yields of the supernovae that enriched successive
stellar generations; one must also deal with the poorly understood
effects of evolutionary mixing that occurs as old stars ascend the red
giant branch.

It is difficult to see how the results for M15 and M92 in
Figure~\ref{Fig:[C/Fe]_Mv}(c), where [C/Fe] decreases as one moves up
the giant branch -- accompanied by increasing [N/Fe] (Langer et al.\
1986, and references therein) -- can be explained in any other way
than by evolutionary mixing involving the CN-cycle (see e.g. Langer et
al.\ 1986), superimposed on the carbon abundance profile of the
protosystems from which these clusters formed.  M15 and M92 have
[Fe/H] $\sim$ --2.2, which lies within $\Delta$[Fe/H] = 0.3~dex of the
mean abundance of {\boo}.  Why then do the Galactic field halo stars
and the {\boo} stars in Figure~\ref{Fig:[C/Fe]_Mv}(a) not share the
extreme evolutionary mixing signature of low carbon on the upper giant
branch that one sees for M15 and M92 (Figure~\ref{Fig:[C/Fe]_Mv}(c))?
One obvious difference that distinguishes the giants of a given
globular cluster is their small age range, which is essentially zero,
while there is no reason to assume this holds for either the Galactic
halo or {\boo}.  That said, it remains unclear why this should produce
the above abundance differences.\footnote{The carbon abundances of the
giants in {\wcen} further complicate the situation.  Persson et al.\
(1980, their Figure 5) first established that this cluster is unique
among Galactic globular clusters in possessing a sub-population of
some 10\% of its red giants that have strong enhancements of their
near infrared CO bands, which is not seen in any other cluster (see
ND, Section~5.3).  As demonstrated by ND (and shown in our
Figure~\ref{Fig:[C/Fe]_Mv}(d), where CO-strong and CO-weak objects are
represented by filled and open circles, respectively) the sub-sample
of CO-strong objects has significantly larger carbon abundances than
that of the CO-weak objects.  We emphasize that no significant
fraction of similar CO-strong objects exists in other clusters.  More
to the point, to our knowledge no satisfactory explanation exists for
this CO-strong sub-population in {\wcen}.}

\subsubsection{Similarities between the [C/Fe] Distributions of the 
Ultra-Faint Dwarf Galaxies and the Galactic Halo}

The above caveats concerning evolutionary mixing effects
notwithstanding, consideration of Figure~\ref{Fig:[C/Fe]_[Fe/H]} and
the upper two panels of Figure~\ref{Fig:[C/Fe]_Mv} suggests a strong
similarity between the carbon abundance distributions of the red
giants of the Galaxy's dwarf galaxies and halo.  At first glance it
might also appear that the available data for the brighter giants in
{\wcen} share this similarity.  Given, however, the biased nature of
the ND sample introduced by their selection towards the chemical
extremes of the cluster, such a conclusion would be premature.  All
one can say is that at least some stars in {\wcen} show abundance
patterns like those of the halo and {\boo}.

If one accepts that the carbon abundance signatures of the Galaxy's
dwarf galaxies and halo are essentially the same, and the conclusion
of Spite et al.\ (2005) that giants with [Fe/H] $\la$ --2.5 comprise
``mixed'' and ``unmixed'' stars, with the ``unmixed'' group
``reflecting the abundances of the early Galaxy'', one might suggest
this could well apply also to the {\uf} dwarfs.  The hypothesis is
testable.  According to Spite et al.\ (2005, see their Figures 6--8)
the ``mixed'' stars have low C and high N (consistent with dredge up
of CNO processed material), together with strong Li dilution, while
the ``unmixed'' objects show no evidence for C to N conversion, and
only moderate Li depletion.  More data to determine nitrogen and
lithium abundances in the dwarf galaxies are clearly needed to address
the question.

Regardless of whether or not the abundances have been modified by
stellar evolutionary effects, the data in the upper panel of
Figure~\ref{Fig:[C/Fe]_[Fe/H]} show that the present relative carbon
abundances [C/Fe] of stars in the Galaxy's {\uf} dwarf satellites are
similar to those of stars in its halo, and in the range --3.5 $<$
[Fe/H] $<$ --2.5 has the value [C/Fe] $\sim$ 0.3.

\subsection{Metallicity Distribution Functions}

\subsubsection{{\boo}}

Accurate iron abundances are now available for 16 radial-velocity
candidate members of {\boo}, and if we restrict this to those within
2.5{\rh} (at which radius, adopting the surface brightness of Martin
et al.\ (2008), the probability of membership is 0.20 of its value at
{\rh}), this number decreases to 15. The outlier, Boo--980, lies at
3.9{\rh} and has [Fe/H] = --3.1, rather suggestive of membership. The
upper panels of Figure~\ref{Fig:MDF} then present the metallicity
distribution function (MDF) (left panel) and the dependence of [Fe/H]
versus elliptical radius (right panel) for all 16 {\boo} stars, while
the lower panels show the results for Galactic dSph systems and
{\wcen} (where, for the latter, radial distance is employed in the
right panel). To improve their formal accuracy, we replaced the {\boo}
AAOmega abundances for Boo--7, 121, and 911 presented in Table~3, for
which we have no high-resolution abundances, with values from Feltzing
et al.\ (2009)\footnote{The [Fe/H] values remain essentially unchanged
(--2.28 vs --2.32), (--2.39, --2.37), and (--2.21, --1.98) for Boo--7,
121, and 911, respectively, while the error bars decrease.}.  There
are three points to note about {\boo}.  First, in comparison with the
steep rise in the MDF seen in {\wcen} between [Fe/H] = --2.2 and
--1.8, its MDF has a slow increase from lowest abundance to the MDF
peak, covering [Fe/H] = --3.7 to --2.3, similar to that seen in the
dSphs, with recent dSph abundance results only accentuating the
low metallicity tails to their MDFs; second, of the dSph systems with
published data for individual stars, it has the highest percentage of
objects with [Fe/H] $<$ --3.0; and third, while there are too few
objects to permit a strong statement, it appears to have a (marginally
significant with these data) radial abundance gradient. Such a
gradient is seen in some of the other dSph systems, as reported by
Winnick (2003), Tolstoy et al.\ (2004), and Koch et al.\ (2006).

\subsubsection{{\seg}}

We noted in Section~1 that Geha et al.\ (2009) provided the first
spectroscopic abundance estimate for {\seg}.  They analyzed a single
radial-velocity member (their star 3451364, with velocity
215.6~{\kms}, 9.5~{\kms} larger than the systemic velocity), and
assuming it to be a red giant reported [Fe/H] = --3.3 $\pm$ 0.2, based
on spectrum synthesis in the CaT wavelength ($\sim$ 8300--8700\ {\AA})
region.  As discussed in Sections 4--6, we have three objects with
reliable abundances and radial velocities consistent with membership
of the {\seg} system, and one other velocity candidate ({\segs}--98)
with inconsistent blue and red abundance estimates.  Of these four
stars, two ({\segs}--31 and {\segs}--71) are relatively metal rich
([Fe/H] $\sim$--2.0), while the C-rich star {\segs}--7 is extremely
metal-poor, with [Fe/H] = --3.5.  These four velocity member stars may
still, statistically, include a chance non-member. It is thus
probable, but not certain, that there exists a large [Fe/H] range in
the {\seg} system.  Clearly, four objects are too few to say much
about the MDF, other than to reiterate our earlier statement that this
system, at face value, is not monometallic.  Additionally, there
remains the open question of whether we are looking at an {\uf} dwarf
(Geha et al.\ 2009), tidal debris from the Sgr dSph (Niederste-Ostholt
et al.\ 2009), or a mixture of both.

\subsubsection{Comparison of metallicity means and dispersions with those of other Milky Way dwarf galaxies}  

Kirby et al.\ (2008) first reported that the mean metallicity of dwarf
galaxies continues to decrease as one proceeds from the more luminous
dSphs to the low absolute magnitudes of the {\uf} systems.  We discuss
here how our results for {\boo} and {\seg} compare with their
relationship, and also investigate the behavior of metallicity
dispersion, $\sigma$([Fe/H]).

In Table~5 we present mean metallicities, $\langle$[Fe/H]$\rangle$,
and dispersions about the mean $\sigma$([Fe/H]) (i.e. standard
deviations) for Milky Way dwarf satellites and the chemically
inhomogeneous globular cluster {\wcen}, together with other data and
their sources.  Columns (1)--(3) contain identification, absolute
visual magnitude M$_{V,\,total}$ and its source, (4)--(5) present mean
metallicity and its error, (6)--(7) the abundance dispersion and its
error, and (8)--(9) give the number of stars in each sample and the
sources of the abundance data. In compiling the table we chose
abundance determinations that met the following criteria: (1) the
observed stars were chosen without chemical bias, which could result,
for example, if selection were biased toward the color extremes in the
CMD in an effort to sample the abundance extremes of the system and
(2) first preference was given to investigations that determined
[Fe/H] from observations that included features of Fe as a primary
observable\footnote{For {\wcen}, Sextans, Carina, UMi, and Draco the
[Fe/H] and $\sigma$([Fe/H])values in Table 3 are based on observations
of the Ca II infrared triplet at 8600\ {\AA}, concerning which we make
the following comments.  First, for {\wcen} Norris and Da Costa (1995)
and Pancino et al.\ (2002) showed that [Ca/Fe] versus [Fe/H] is
well-determined and follows the Galactic halo relationship, permitting
one to transform the published values of [Ca/H] to [Fe/H] using the
relationship of the Galactic halo.  Second, concerning the use of the
Ca triplet to determine [Fe/H] for Sextans, Carina, UMi, and Draco we
recall that the data of Battaglia et al.\ (2008), obtained for Fornax
and Sculptor using the same technique, show that the Ca II triplet and
higher resolution spectroscopy yield the same values of
$\sigma$([Fe/H]), to within 0.04 -- somewhat surprisingly to us, given
their non-Galactic halo [Ca/Fe] versus [Fe/H] relationships.}. In
determining uncertainties in the abundance dispersions, we followed
the precepts of Da Costa et al.\ (1977), which explicitly remove the
effects of instrumental and statistical abundance errors in the
determination of dispersion.  For {\boo}, we used the data in Table~3,
to which we assign measurement errors of 0.15 dex for the ten
high-resolution analyses (including the three Feltzing et al.\ (2009)
stars discussed in Section 7.1), on the one hand, and 0.35 dex for the
six remaining AAOmega-based result, on the other, while for the {\seg}
objects we adopted $\Delta$[Fe/H] = 0.4 dex for {\segs}--31 and
{\segs}--71) and 0.2 dex for the Geha et al. red giant and {\segs}--7.
For the other systems we assumed abundance errors of 0.2 dex for the
{\uf} dwarfs, UMi, and Draco (based on error estimates in Kirby et
al.\ (2008) and data of Winnick (2003)), and 0.10 dex for the other
dSphs and {\wcen} (following Battaglia et al.\ (2008), Helmi et al.\
(2006), and Norris et al.\ (1996)).

Figure~\ref{Fig:SigFe_vs_Mv} presents mean metallicities and
dispersions about the mean as a function of integrated absolute visual
magnitude M$_{V,\,total}$.  Inspection of the figure shows the
continued decrease in mean [Fe/H] as one proceeds to lower absolute
magnitude, first reported by Kirby et al.\ (2008), together with an
apparent increase in the abundance dispersion. Whether or not this
trend continues to the luminosity of {\seg} is uncertain, being
sensitive to membership assignments. The two {\seg} stars
with [Fe/H] $\sim$ --2. may be field interlopers. The two very
metal-poor stars are offset in either velocity ({\segs}--3451364) or
position ({\segs}--7) from the reported systemic values.  More data are
clearly required for {\seg} to test this suggestion more rigorously.
That said, an increased abundance spread is intuitively consistent
with inhomogeneous stochastic enrichment in very low-baryonic-mass and
low-metallicity systems, enriched by very few supernova events.

\subsection{The Carbon-Rich, Extremely-Metal-Poor Star in {\seg}}

{\segs}--7 has a spectrum that is similar to those of the C-rich,
extremely-metal poor stars of the Galactic halo.  As emphasized above,
more work is needed to investigate the possible connection between the
parent populations of the Galaxy's dwarf satellites and halo.  The
question is an important one for an understanding of chemical
enrichment at the very earliest times.  As one proceeds to lowest
abundance in the Galactic halo (i.e. to the oldest stars) one finds
that the fraction of C-rich objects increases -- indeed all stars to
date with [Fe/H] $\la$ --4.3 are C-rich (see e.g. Norris et al.\
2007).  Further, efforts to understand the intriguing relative
abundance distributions of C-rich stars with [Fe/H] $\la$ --3.5, many
of which are characterized by enhanced values of some, or all, of
[C/Fe], [N/Fe], [O/Fe], and [Mg/Fe] of order 1--2 dex, have led to the
proposal of severely non-canonical supernovae at the earliest times
and lowest metallicities, involving for example ``mixing and
fallback'' during explosion (see Iwamoto et al.\ 2005, and references
therein).  While it has been proposed that the elemental abundances of
the most metal-poor stars in the field halo can be explained by the
explosions of zero-metallicity massive stars with a Salpeter IMF over
the range 15--40M$_\odot$ (Joggerst et al.\ 2010), we note that these
objects do not produce the extreme CNO abundances of a significant
fraction of CEMP stars with [Fe/H] $<$ --3.5. If more exotic
supernovae are a natural feature of the evolution of extremely
metal-poor massive stars, they may have also left their signatures in
the lowest abundance stars within the {\uf} systems.

\section {DISCUSSION AND SUMMARY}

The primary conclusion of our investigation is that the spread of
carbon abundance in both {\boo} and {\seg} is large.  For {\boo}, four
of the 16 objects in Table~3 have [C/H] $\la$ --3.1, while two have
[C/H] $\ga$ --2.3, suggesting a range of $\Delta$[C/H] $\sim$
0.8. In Table~4 one finds that for {\seg} there exists a range
$\Delta$[C/H]~$\sim$~1.0. (For iron, the accompanying ranges are
$\Delta$[Fe/H]~$\sim$~1.6 for both {\boo} and {\seg}.) A related
result is that for [Fe/H] $<$ --3.0 the Milky Way's dwarf satellites
exhibit a dependence of [C/Fe] on [Fe/H] that is very similar to that
observed for the Galaxy's halo.  We find [C/Fe]~$\sim$ 0.3 for stars
in the {\uf} systems that we believe are the counterpart of the Spite
et al. (2005) ``unmixed'' giants of the Galactic halo and for which
they report [C/Fe]~$\sim$~0.2.  Of particlar interest is that the
{\seg} system contains an extremely metal-poor, carbon-rich star, with
[Fe/H] = --3.5, [C/Fe] = +2.3.

The second conclusion is confirmation of the correlation between
(decreasing) luminosity and (decreasing) mean metallicity in dwarf
galaxies, Figure~\ref{Fig:SigFe_vs_Mv}.  This luminosity-metallicity
correlation is consistent with models of a mass-metallicity relation
in which system mass (predominantly dark matter) provides the
potential well to (at least partially) retain supernova ejecta, while
the supernova rate is never so high that it evacuates all the
remaining gas before further star formation can happen -- that is,
self-limiting star formation with steady gas loss.  It is also
consistent with models where there is no system mass-metallicity
relation, where all (dark matter) potential wells are similar, and the
wide range of surviving luminosities depends on a gas-loss rate driven
by very different star formation rates, in the extreme cases implying
a very short duration of star formation before system gas
loss. Consideration of the dispersion in stellar abundances and in
stellar abundance ratios is necessary to quantify the efficiency of
mixing of enrichment, and the number of supernovae whose ejecta were
retained during continuing star formation. The first model is
consistent with stellar abundances reflecting enrichment by the mixed
ejecta of many supernovae, the second predicts substantial star to
star scatter in enrichment patterns.

More fundamentally, the existence of this relation shows that the
observed dwarfs are not drastically stripped remnants of initially
more luminous systems, which by the luminosity-metallicity relation,
would have been more metal-rich. This robust conclusion is surprising,
as generic hierarchical galaxy formation models require that the few
dwarf galaxies that are currently found within the inner halo of the
Milky Way should be short-lived debris near final disruption, from an
initally much more luminous state.

The continuation of this relation to luminosities below that of {\boo}
remains to be proven. The available data are equally consistent with a
correlation continuing down to {\seg}, at M$_V$ = --2, and a correlation
flattening at the luminosity of {\boo}, M$_V$ = --6.

The other clear feature of the metallicity distribution functions
shown in Figure~\ref{Fig:MDF} is the special status of {\wcen}, which
is unique in not having a tail of stars to very low metallicity.  The
simplest explanation of this is that it formed from gas that had been
substantially pre-enriched prior to its own formation, which is not
the case for any known dSph galaxy.  This argues strongly against the
hypothesis that {\wcen} is the surviving remnant/core of the bulk of a
dSph parent, but rather argues that {\wcen} is a luminous star
cluster.  (A similar situation may also pertain to the globular
cluster M54 and the bulk of the disintegrating Sgr dSph.)  The
essential difference between dSphs and {\wcen} is most likely that for
the former it was primarily dark matter that facilitated self
enrichment from material having very low initial abundance, while
proto-{\wcen} contained sufficient baryonic mass to retain some
stellar ejecta and so host later generations of star formation
(perhaps within a larger parent containing dark matter which had
experienced some chemical evolution and provided the pre-enrichment).

\subsection{What is {\seg}?}

The nature of {\seg} remains the subject of active discussion: while
Geha et al.\ (2009) identify it as an {\uf} dwarf galaxy,
Niederste-Ostholt et al.\ (2009) suggest that it is probably a
disrupted system, perhaps initially a star cluster, associated with
the Sagittarius galaxy.  This latter identification was made on the
basis of the complex system morphology, combined with the likelihood
that interlopers from the Sgr stream both contaminate the abundance
distribution function, and, if present, would inflate the derived
velocity dispersion of {\seg} and contribute to an (erroneous)
identification as an {\uf} galaxy.  In this study we do not contribute
to the kinematic analysis usefully. What of our abundances? Our
results are summarized in Table 4, and discussed above.

We have two radial-velocity candidate members with [Fe/H] $\sim$ $-2.$,
and two with [Fe/H] $\sim$ $-3.5$, one of which has a somewhat offset
velocity (by $\sim$10~{\kms} from the systemic value), the other with
a somewhat offset spatial position (by 17{\arcmin} from galaxy
center). We must allow for the possibility of significant
contamination of our sample.

This unsatisfactory data set allows three possible options.  (1) At
least three of the four stars are system members. In that case {\seg}
indeed has a broad range of iron abundances, comparable to that seen
in the more luminous dSphs, and a ``mean'' abundance, insofar as a
patently bimodal distribution function can be so characterized,
implying that {\seg} is (or was) embedded in a sufficiently massive
(dark) halo to retain supernova ejecta, and self-enrich. That is to
say, {\seg} was, and may still be, an {\uf} galaxy; (2) The two most
metal-poor stars, including the carbon-rich star, are contaminants
from the Sgr dwarf. In this case {\seg} has no detected metallicity
dispersion, has mean metallicity similar to that of other Sgr globular
clusters, and {\seg} is another Sgr star cluster. The two extremely
metal-poor stars then require separate explanation. (3) The two most
metal-rich stars, which have abundances similar to those of member
stars in the Sgr dwarf, are interlopers, with abundances similar to
those of other Sgr debris.  In this case {\seg} is an extremely
metal-poor star cluster embedded in Sgr debris.

Each possibility is interesting.  Statistically, given the rarity of
extremely metal-poor carbon-enhanced stars, we slightly favor option
(1), that {\seg} is indeed an exceptionally low-luminosity galaxy,
whose abundances suggest it is the lowest luminosity, most primitive
dwarf galaxy yet discovered.  (To us, possibility (2), that both
extremely metal-poor stars are interlopers from a relatively
metal-richer Sgr seems unlikely, given their rarity, while (3), that
we are dealing with a star cluster of very low mass
($\sim$~1000{\msun}, Martin et al. 2008), raises questions concerning
the survival of the system against disruption.) The apparent bimodal
abundance distribution, if it survives larger data sets, suggests a
very low star formation rate at early times, perhaps even with a pause
after formation of the early stars with [Fe/H] $\sim$
--3.5 (including the carbon-rich object) and/or extremely
inhomogeneous mixing of ejecta in the local ISM. The implied low star
formation rate is also consistent with survival of {\seg}
as a system with continuing gas retention. It is probable that {\seg}
is associated with Sgr, rather than that it is an isolated system, so
that {\seg} is a sub-halo of Sgr, and hence has recently become a
sub-sub-halo of the Milky Way. The extremely low representative
metallicity of {\seg} is consistent with it being an extremely early
system, with its earliest surviving stars showing an element pattern
suggesting enrichment by the first generation of near zero-metal
supernovae. The consistency of {\seg} with a continuous
luminosity-metallicity relation argues strongly that {\seg} has always
been an ultra low-luminosity system. It is not tidal debris from an
initially much more luminous system. Thus {\seg} becomes a candidate
survivor from the dark ages, pre-reionization.

\acknowledgements

The authors gratefully acknowledge the contributions of the AAOmega
project team, in particular Rob Sharp, during this investigation.  We
also thank Evan Kirby and Joshua Simon for discussions on their
ongoing investigation of {\seg}.  Studies at RSAA, ANU, of the most
metal-poor stellar populations are supported by Australian Research
Council grants DP0663562 and DP0984924, which J.E.N. and D.Y. are
pleased to acknowledge.  R.F.G.W. acknowledges grants from the
W.M. Keck Foundation and the Gordon \& Betty Moore Foundation, to
establish a program of data-intensive science at the Johns Hopkins
University.  She also thanks all at the Institute for Astronomy,
Edinburgh University for their hospitality during her appointment as
Distinguished Visitor.  A.F. is supported by a Clay Fellowship
administered by the Smithsonian Astrophysical Observatory.

\noindent{\it Facilities:} {AAT(AAOmega), ATT(DBS)}

\clearpage

\begin{deluxetable}{rccrrrrrrr}
\tablecaption{OBSERVATIONAL DATA AND DERIVED HELIOCENTRIC VELOCITIES FOR STARS IN THE BO{\"O}TES I FIELD\vspace{0mm}}
\tablewidth{0pt}
\tablehead{
\colhead {ID} & {${\rm RA}_{2000}$}           & ${\rm Dec}_{2000}$        & {g} & {i}  & {X}  & {Y} & {R} & {$v_{helio}$} & {Candidate} \\

& & & & &  {(\arcmin)} & {(\arcmin)} & {(\arcmin)} & {(\kms)} &  {member?}}
\startdata
2 & 13 59 2.33 & +14 30 23.2 & 18.33 & 17.34 & $-$15.4 & 0.4 & 15.4 & 123	& y \\
3 & 13 58 59.98 & +14 23 40.5 & 18.31 & 17.21 & $-$16.0 & $-$6.3 & 17.2 & $-$38	& n \\
7 & 13 59 35.53 & +14 20 23.7 & 18.31 & 17.29 & $-$7.4 & $-$9.6 & 12.1 & 106 &	y   \\
8 & 13 59 38.62 & +14 19 15.9 & 19.03 & 18.20 & $-$6.6 & $-$10.7 & 12.6 & 106	& y \\
9 & 13 59 48.81 & +14 19 42.9 & 17.92 & 16.75 & $-$4.2 & $-$10.3 & 11.1 & 112	& y \\
23 & 14 00 1.54 & +14 21 54.2 & 19.89 & 19.06 & $-$1.1 & $-$8.1 & 8.2 & 105	& y  \\
32 & 14 00 7.06 & +14 20 27.0 & 20.08 & 19.28 & 0.3 & $-$9.6 & 9.6 & 105	& y  \\
33 & 14 00 11.73 & +14 25 1.4 & 18.23 & 17.16 & 1.4 & $-$5.0 & 5.2 & 107	& y \\
34 & 14 00 21.11 & +14 17 27.9 & 18.74 & 17.75 & 3.7 & $-$12.5 & 13.1 & 111	& y \\
41 & 14 00 25.83 & +14 26 7.6 & 18.38 & 17.37 & 4.8 & $-$3.9 & 6.2 & 105 &   y\\
45 & 14 00 26.18 & +14 27 29.5 & 19.08 & 18.24 & 4.9 & $-$2.5 & 5.5 & 65	& n \\
53 & 14 00 40.78 & +14 17 22.0 & 20.06 & 19.30 & 8.4 & $-$12.6 & 15.2 & 14	& n \\
58 & 13 59 20.47 & +14 44 56.9 & 19.44 & 18.62 & $-$11.0 & 15.0 & 18.6 & 30	& n \\
63 & 14 00 32.00 & +14 43 2.5 & 18.22 & 17.13 & 6.3 & 13.0 & 14.5 & 97	& y \\
66 & 14 00 40.38 & +14 45 44.9 & 18.17 & 17.15 & 8.3 & 15.8 & 17.8 & $-$76	& n \\
71 & 13 59 36.79 & +14 11 40.8 & 19.59 & 18.84 & $-$7.1 & $-$18.3 & 19.6 & 91	& y \\
73 & 13 59 36.20 & +14 14 52.4 & 19.22 & 18.42 & $-$7.2 & $-$15.1 & 16.8 & $-$3	& n \\
74 & 13 59 43.41 & +14 16 1.6 & 18.83 & 18.00 & $-$5.5 & $-$14.0 & 15.0 & 27	& n \\
76 & 14 00 3.02 & +14 12 36.0 & 20.08 & 19.23 & $-$0.7 & $-$17.4 & 17.4 & 95	& y \\
77 & 14 00 28.55 & +14 13 44.4 & 18.43 & 17.41 & 5.5 & $-$16.3 & 17.2 & $-$12	& n \\
78 & 14 00 14.73 & +14 13 13.9 & 19.30 & 18.36 & 2.1 & $-$16.8 & 16.9 & 105	& y \\
83 & 13 58 50.77 & +14 35 30.4 & 19.94 & 19.31 & $-$18.2 & 5.5 & 19.0 & 16	& n \\
84 & 13 58 55.98 & +14 40 35.1 & 18.70 & 17.72 & $-$16.9 & 10.6 & 20.0 & $-$38	& n \\
85 & 13 59 15.34 & +14 35 52.8 & 18.85 & 17.90 & $-$12.3 & 5.9 & 13.6 & $-$4	& n \\
91 & 13 59 35.66 & +14 37 35.0 & 19.97 & 19.14 & $-$7.3 & 7.6 & 10.6 & 2	& n \\
92 & 13 59 21.37 & +14 36 6.3 & 17.83 & 16.68 & $-$10.8 & 6.1 & 12.4 & $-$14	& n \\
94 & 14 00 31.51 & +14 34 3.6 & 17.53 & 16.26 & 6.2 & 4.1 & 7.4 & 94	& y \\
98 & 13 59 8.12 & +14 32 28.1 & 18.58 & 17.59 & $-$14.0 & 2.5 & 14.2 & 20	& n \\
105 & 14 00 12.92 & +14 33 11.8 & 19.54 & 18.72 & 1.7 & 3.2 & 3.6 & 98	& y \\
111 & 14 00 22.97 & +14 30 45.0 & 18.55 & 17.50 & 4.1 & 0.8 & 4.2 & $-$45	& n \\
117 & 14 00 10.49 & +14 31 45.5 & 18.19 & 17.13 & 1.1 & 1.8 & 2.1 & 96	& y \\
121 & 14 00 36.52 & +14 39 27.3 & 17.92 & 16.72 & 7.4 & 9.5 & 12.0 & 112	& y \\
127 & 14 00 14.57 & +14 35 52.7 & 18.15 & 17.00 & 2.1 & 5.9 & 6.2 & 99	& y \\
130 & 13 59 48.98 & +14 30 6.2 & 18.21 & 17.15 & $-$4.1 & 0.1 & 4.1 & 111	& y \\
161 & 13 59 39.37 & +14 26 38.4 & 20.38 & 19.67 & $-$6.4 & $-$3.4 & 7.3 & 127	& y \\
162 & 13 59 46.34 & +14 25 11.8 & 20.36 & 19.66 & $-$4.8 & $-$4.8 & 6.8 & 91	& y \\
203 & 13 58 52.92 & +14 25 5.3 & 20.78 & 20.00 & $-$17.7 & $-$4.9 & 18.4 & $-$24	& n \\
229 & 13 59 29.95 & +14 14 56.7 & 20.57 & 19.83 & $-$8.7 & $-$15.1 & 17.4 & 126	& y \\
237 & 14 00 42.15 & +14 15 20.4 & 20.51 & 19.83 & 8.8 & $-$14.7 & 17.1 & 5	& n \\
244 & 13 59 9.10 & +14 33 11.3 & 20.69 & 19.93 & $-$13.8 & 3.2 & 14.1 & 113	& y \\
306 & 14 03 52.07 & +14 14 33.5 & 19.02 & 18.06 & 54.8 & $-$15.3 & 56.9 & 123	& y \\
337 & 14 03 27.88 & +14 15 10.3 & 20.23 & 19.51 & 48.9 & $-$14.7 & 51.1 & $-$144	& n \\
342 & 14 03 45.02 & +14 48 33.9 & 20.07 & 19.34 & 52.9 & 18.7 & 56.1 & $-$162	& n \\
343 & 14 03 25.77 & +14 23 6.2 & 17.76 & 16.54 & 48.4 & $-$6.8 & 48.9 & 11	& n \\
354 & 14 03 18.84 & +14 18 25.6 & 18.00 & 16.83 & 46.7 & $-$11.5 & 48.1 & $-$39	& n \\
357 & 14 03 3.33 & +13 57 8.8 & 19.50 & 18.78 & 43.0 & $-$32.8 & 54.1 & 118	& y \\
399 & 14 03 5.58 & +14 31 19.5 & 18.90 & 17.90 & 43.5 & 1.4 & 43.5 & $-$39	& n \\
414 & 14 02 41.93 & +14 04 56.4 & 18.15 & 17.02 & 37.8 & $-$25.0 & 45.3 & $-$12	& n \\
425 & 14 02 53.39 & +14 31 59.3 & 20.02 & 19.33 & 40.5 & 2.0 & 40.6 & 171	& n \\
436 & 14 02 24.00 & +13 51 16.2 & 19.97 & 19.20 & 33.5 & $-$38.7 & 51.2 & 26	& n \\
463 & 14 02 14.20 & +13 49 54.4 & 18.54 & 17.64 & 31.1 & $-$40.1 & 50.7 & 63	& n \\
465 & 14 02 23.66 & +14 09 28.6 & 19.31 & 18.51 & 33.4 & $-$20.5 & 39.2 & $-$6	& n \\
484 & 14 02 27.15 & +14 23 13.0 & 20.36 & 19.61 & 34.2 & $-$6.7 & 34.8 & 45	& n \\
492 & 14 02 19.02 & +14 18 25.3 & 18.42 & 17.29 & 32.2 & $-$11.5 & 34.2 & $-$12	& n \\
528 & 14 01 39.94 & +13 35 55.1 & 18.59 & 17.58 & 22.8 & $-$54.1 & 58.7 & $-$21	& n \\
572 & 14 01 30.41 & +13 48 56.6 & 19.13 & 18.22 & 20.5 & $-$41.0 & 45.9 & 75	& n \\
574 & 14 01 41.91 & +14 08 43.0 & 19.66 & 18.95 & 23.3 & $-$21.3 & 31.5 & $-$145	& n \\
582 & 14 01 42.53 & +14 15 20.1 & 18.36 & 17.39 & 23.4 & $-$14.6 & 27.6 & 3	& n \\
598 & 14 01 17.19 & +13 38 43.1 & 20.62 & 19.87 & 17.3 & $-$51.3 & 54.1 & $-$113	& n \\
625 & 14 01 45.40 & +14 41 54.7 & 19.86 & 19.15 & 24.0 & 11.9 & 26.8 & 85	& y \\
634 & 14 01 5.84 & +13 38 46.4 & 18.37 & 17.35 & 14.5 & $-$51.2 & 53.2 & $-$37	& n \\
649 & 14 01 27.35 & +14 24 43.9 & 20.20 & 19.54 & 19.7 & $-$5.3 & 20.4 & 125	& y \\
653 & 14 01 22.38 & +14 17 24.3 & 18.31 & 17.25 & 18.5 & $-$12.6 & 22.4 & 29	& n \\
655 & 14 01 40.25 & +14 48 20.2 & 18.63 & 17.69 & 22.8 & 18.4 & 29.3 & 0	& n \\
659 & 14 01 51.99 & +15 10 55.0 & 19.96 & 19.28 & 25.6 & 40.9 & 48.3 & $-$71	& n \\
684 & 14 01 50.19 & +15 20 6.1 & 20.75 & 20.12 & 25.1 & 50.1 & 56.1 & 32	& n \\
688 & 14 00 50.29 & +13 45 32.4 & 20.56 & 19.78 & 10.8 & $-$44.5 & 45.7 & 69	& n \\
690 & 14 01 5.34 & +14 11 56.6 & 20.19 & 19.46 & 14.4 & $-$18.0 & 23.1 & $-$47	& n \\
693 & 14 01 32.48 & +14 57 23.3 & 18.79 & 17.82 & 20.9 & 27.4 & 34.5 & $-$140	& n \\
701 & 14 00 56.44 & +14 02 53.3 & 20.06 & 19.38 & 12.2 & $-$27.1 & 29.7 & 75	& n \\
706 & 14 00 49.47 & +13 54 19.6 & 19.74 & 18.87 & 10.6 & $-$35.7 & 37.2 & $-$18	& n \\
720 & 14 00 30.65 & +13 32 18.8 & 18.67 & 17.60 & 6.0 & $-$57.7 & 58.0 & $-$32	& n \\
738 & 14 00 37.22 & +13 56 16.3 & 20.25 & 19.45 & 7.6 & $-$33.7 & 34.6 & 65	& n \\
774 & 14 00 17.08 & +13 39 32.8 & 19.36 & 18.47 & 2.7 & $-$50.5 & 50.5 & 98	& y \\
815 & 14 00 51.41 & +15 00 41.7 & 18.62 & 17.68 & 11.0 & 30.7 & 32.6 & $-$42	& n \\
817 & 14 00 17.29 & +14 09 36.5 & 20.30 & 19.65 & 2.7 & $-$20.4 & 20.6 & 40	& n \\
818 & 13 59 55.45 & +13 31 43.8 & 20.65 & 19.93 & $-$2.6 & $-$58.3 & 58.3 & $-$207	& n \\
822 & 14 00 9.49 & +13 58 21.3 & 19.30 & 18.46 & 0.8 & $-$31.6 & 31.7 & $-$21	& n \\
856 & 13 59 49.55 & +13 44 29.9 & 17.91 & 16.85 & $-$4.0 & $-$45.5 & 45.7 & 17	& n \\
859 & 14 00 6.47 & +14 16 30.5 & 19.89 & 19.10 & 0.1 & $-$13.5 & 13.5 & 114	& y \\
870 & 14 00 37.27 & +15 12 18.2 & 17.81 & 16.73 & 7.5 & 42.3 & 43.0 & $-$8	& n \\
876 & 13 59 38.53 & +13 38 37.1 & 18.53 & 17.55 & $-$6.7 & $-$51.4 & 51.8 & $-$35	& n \\
911 & 14 00 1.08 & +14 36 51.5 & 17.94 & 16.82 & $-$1.2 & 6.9 & 7.0 & 94	& y \\
912 & 13 59 39.07 & +14 00 14.6 & 20.94 & 20.31 & $-$6.5 & $-$29.8 & 30.5 & $-$27	& n \\
923 & 13 59 53.76 & +14 30 56.0 & 20.65 & 19.88 & $-$3.0 & 0.9 & 3.1 & 115	& y \\
952 & 13 59 30.37 & +14 02 51.2 & 20.61 & 20.00 & $-$8.6 & $-$27.1 & 28.5 & 70	& n \\
980 & 13 59 12.68 & +13 42 55.8 & 18.51 & 17.60 & $-$12.9 & $-$47.1 & 48.8 & 106	& y \\
986 & 13 59 19.21 & +13 56 43.0 & 19.04 & 18.12 & $-$11.4 & $-$33.3 & 35.2 & $-$30	& n \\
992 & 13 59 15.22 & +13 52 11.8 & 20.03 & 19.29 & $-$12.3 & $-$37.8 & 39.8 & 68	& n \\
1004 & 13 59 20.47 & +14 08 20.4 & 18.72 & 17.71 & $-$11.0 & $-$21.7 & 24.3 & $-$55	& n \\
1049 & 13 59 37.88 & +15 11 39.5 & 19.24 & 18.35 & $-$6.8 & 41.7 & 42.2 & $-$20	& n \\
1051 & 13 58 44.21 & +13 41 5.8 & 19.49 & 18.76 & $-$19.9 & $-$48.9 & 52.8 & $-$56	& n \\
1069 & 13 58 53.22 & +14 06 57.8 & 19.05 & 18.23 & $-$17.6 & $-$23.0 & 29.0 & 113	& y \\
1082 & 13 58 34.41 & +13 42 7.5 & 20.22 & 19.54 & $-$22.2 & $-$47.9 & 52.8 & 0	& n \\
1090 & 13 58 33.05 & +13 49 17.3 & 20.27 & 19.61 & $-$22.6 & $-$40.7 & 46.5 & $-$95	& n \\
1124 & 13 58 23.90 & +13 55 14.8 & 17.89 & 16.67 & $-$24.8 & $-$34.7 & 42.7 & $-$42	& n \\
1126 & 13 58 16.42 & +13 42 49.4 & 20.74 & 20.05 & $-$26.6 & $-$47.1 & 54.1 & $-$52	& n \\
1133 & 13 58 37.31 & +14 23 26.0 & 20.44 & 19.75 & $-$21.5 & $-$6.5 & 22.5 & $-$121	& n \\
1134 & 13 58 23.81 & +14 01 19.5 & 18.45 & 17.44 & $-$24.8 & $-$28.7 & 37.9 & $-$47	& n \\
1137 & 13 58 33.82 & +14 21 8.5 & 18.11 & 17.04 & $-$22.3 & $-$8.8 & 24.0 & 108	& y \\
1177 & 13 58 5.96 & +13 56 19.6 & 18.79 & 17.90 & $-$29.1 & $-$33.6 & 44.5 & $-$44	& n \\
1182 & 13 58 20.67 & +14 25 3.7 & 20.39 & 19.73 & $-$25.5 & $-$4.9 & 26.0 & 327	& n \\
1186 & 13 58 32.07 & +14 46 3.1 & 18.93 & 17.96 & $-$22.7 & 16.1 & 27.8 & $-$15	& n \\
1204 & 13 58 35.65 & +15 05 4.8 & 20.52 & 19.79 & $-$21.8 & 35.1 & 41.3 & 108	& y \\
1207 & 13 58 1.58 & +14 09 25.0 & 18.31 & 17.28 & $-$30.2 & $-$20.5 & 36.5 & $-$113	& n \\
1225 & 13 57 41.79 & +13 47 16.5 & 20.56 & 19.78 & $-$35.0 & $-$42.7 & 55.2 & 2	& n \\
1254 & 13 57 40.05 & +14 02 4.8 & 20.55 & 19.91 & $-$35.4 & $-$27.9 & 45.1 & $-$22	& n \\
1259 & 13 57 35.61 & +13 57 4.9 & 19.30 & 18.39 & $-$36.5 & $-$32.9 & 49.1 & $-$46	& n \\
1260 & 13 57 42.24 & +14 09 43.9 & 19.44 & 18.54 & $-$34.8 & $-$20.2 & 40.3 & $-$78	& n \\
1264 & 13 57 42.11 & +14 12 27.8 & 18.49 & 17.39 & $-$34.9 & $-$17.5 & 39.0 & $-$11	& n \\
1271 & 13 58 10.50 & +15 04 50.4 & 18.56 & 17.59 & $-$27.9 & 34.9 & 44.6 & $-$52	& n \\
1297 & 13 57 32.72 & +14 15 49.1 & 20.32 & 19.54 & $-$37.1 & $-$14.1 & 39.7 & $-$115	& n \\
1313 & 13 57 28.72 & +14 18 30.3 & 19.73 & 18.96 & $-$38.1 & $-$11.4 & 39.8 & 59	& n \\
1314 & 13 57 27.50 & +14 17 2.8 & 20.51 & 19.80 & $-$38.4 & $-$12.9 & 40.5 & 103	& y \\
1349 & 13 57 2.69 & +13 59 18.3 & 19.05 & 18.17 & $-$44.5 & $-$30.6 & 54.0 & $-$100	& n \\
1378 & 13 56 50.62 & +13 54 2.0 & 20.50 & 19.74 & $-$47.4 & $-$35.9 & 59.5 & 40	& n \\
1391 & 13 56 49.09 & +13 59 8.2 & 18.74 & 17.77 & $-$47.8 & $-$30.8 & 56.8 & $-$93	& n \\
1439 & 13 56 50.63 & +14 31 58.7 & 19.50 & 18.65 & $-$47.3 & 2.1 & 47.3 & 17	& n \\
1449 & 13 56 35.02 & +14 11 13.3 & 18.29 & 17.28 & $-$51.1 & $-$18.7 & 54.4 & $-$34	& n \\
1462 & 13 56 36.17 & +14 24 15.0 & 20.22 & 19.46 & $-$50.8 & $-$5.7 & 51.1 & $-$34	& n \\
1473 & 13 56 19.22 & +14 05 49.2 & 19.61 & 18.82 & $-$55.0 & $-$24.1 & 60.0 & 19	& n \\
1483 & 13 56 19.44 & +14 22 25.6 & 18.76 & 17.81 & $-$54.9 & $-$7.5 & 55.4 & 35	& n \\
\enddata
\end{deluxetable}

\clearpage

\begin{deluxetable}{rccrrrrrrc}
\tablecaption{OBSERVATIONAL DATA AND DERIVED HELIOCENTRIC VELOCITIES FOR STARS IN THE SEGUE 1 FIELD\vspace{0mm}}
\tablewidth{0pt}
\tablehead{
\colhead {ID} & {${\rm RA}_{2000}$}           & ${\rm Dec}_{2000}$      & {g} & {i}  & {X}  & {Y} & {R} & {$v_{helio}$} & {Candidate} \\

& & & & &  {(\arcmin)} & {(\arcmin)} & {(\arcmin)} & {(\kms)} & {member?} }
\startdata
1 & 10 9 49.84 & 16 05 9.1 & 18.21 & 17.57 & 39.8 & 0.3 & 39.8 & $-$67	& n \\
6 & 10 9 48.59 & 16 03 52.6 & 18.73 & 18.09 & 39.5 & $-$1.0 & 39.6 & 9	& n \\
7 & 10 8 14.45 & 16 05 1.2 & 18.05 & 17.30 & 16.9 & 0.1 & 16.9 & 194	& y \\
11 & 10 9 14.95 & 15 59 48.4 & 17.97 & 17.25 & 31.5 & $-$5.1 & 31.9 & 301	& n \\
15 & 10 7 53.60 & 16 00 33.6 & 18.90 & 18.25 & 11.9 & $-$4.4 & 12.7 & 76	& n \\
16 & 10 7 20.53 & 16 03 16.1 & 17.67 & 16.87 & 4.0 & $-$1.6 & 4.3 & $-$32	& n \\
21 & 10 8 18.89 & 15 57 50.1 & 17.85 & 17.09 & 18.0 & $-$7.1 & 19.3 & 50	& n \\
22 & 10 7 26.19 & 16 03 54.3 & 17.47 & 16.50 & 5.3 & $-$1.0 & 5.4 & $-$4	& n \\
23 & 10 7 40.13 & 16 03 9.7 & 19.95 & 19.43 & 8.7 & $-$1.8 & 8.9 & 298	& n \\
24 & 10 7 52.01 & 15 56 23.0 & 17.25 & 16.53 & 11.5 & $-$8.5 & 14.3 & 48	& n \\
28 & 10 7 55.19 & 15 51 23.7 & 20.28 & 19.58 & 12.3 & $-$13.5 & 18.3 & $-$17	& n \\
31 & 10 7 42.72 & 16 01 6.9 & 18.59 & 17.91 & 9.3 & $-$3.8 & 10.0 & 211	& y \\
34 & 10 8 15.24 & 15 49 19.6 & 19.71 & 19.18 & 17.1 & $-$15.6 & 23.2 & $-$11	& n \\
36 & 10 8 31.87 & 15 55 23.9 & 19.72 & 19.06 & 21.1 & $-$9.5 & 23.2 & 167	& n \\
38 & 10 7 46.02 & 15 55 24.0 & 18.80 & 17.99 & 10.1 & $-$9.5 & 13.9 & $-$128	& n \\
39 & 10 7 52.76 & 15 57 39.0 & 20.26 & 19.75 & 11.7 & $-$7.3 & 13.8 & 0	& n \\
39 & 10 7 52.76 & 15 57 39.0 & 20.26 & 19.75 & 11.7 & $-$7.3 & 13.8 & 9	& n \\
40 & 10 8 36.49 & 15 55 40.3 & 17.59 & 16.65 & 22.2 & $-$9.2 & 24.1 & 81	& n \\
42 & 10 7 33.00 & 15 58 34.6 & 18.73 & 18.08 & 7.0 & $-$6.3 & 9.4 & 195	& y \\
44 & 10 7 38.95 & 15 58 15.3 & 18.30 & 17.40 & 8.4 & $-$6.7 & 10.7 & 30	& n \\
46 & 10 7 58.23 & 15 55 25.5 & 18.28 & 17.56 & 13.0 & $-$9.5 & 16.1 & $-$2	& n \\
54 & 10 7 23.72 & 15 54 7.5 & 19.71 & 19.18 & 4.7 & $-$10.8 & 11.8 & 108	& n \\
60 & 10 7 37.57 & 15 45 18.9 & 18.41 & 17.62 & 8.1 & $-$19.6 & 21.2 & 47	& n \\
63 & 10 8 52.37 & 15 38 29.7 & 19.02 & 18.28 & 26.1 & $-$26.4 & 37.1 & 49	& n \\
68 & 10 7 22.54 & 15 53 33.5 & 20.00 & 19.45 & 4.5 & $-$11.4 & 12.2 & $-$9	& n \\
68 & 10 7 22.54 & 15 53 33.5 & 20.00 & 19.45 & 4.5 & $-$11.4 & 12.2 & $-$9	& n \\
71 & 10 7 2.46 & 15 50 55.3 & 18.46 & 17.72 & $-$0.4 & $-$14.0 & 14.0 & 212	& y \\
74 & 10 7 28.04 & 15 40 23.2 & 20.26 & 19.75 & 5.8 & $-$24.5 & 25.2 & 10	& n \\
76 & 10 7 41.99 & 15 54 56.2 & 19.63 & 19.09 & 9.1 & $-$10.0 & 13.5 & 24	& n \\
78 & 10 7 28.05 & 15 45 53.3 & 20.05 & 19.29 & 5.8 & $-$19.0 & 19.9 & $-$54	& n \\
84 & 10 7 31.49 & 15 46 13.5 & 20.68 & 20.20 & 6.6 & $-$18.7 & 19.8 & 44	& n \\
84 & 10 7 31.49 & 15 46 13.5 & 20.68 & 20.20 & 6.6 & $-$18.7 & 19.8 & 57	& n \\
86 & 10 7 9.01 & 15 43 27.3 & 19.62 & 18.86 & 1.2 & $-$21.5 & 21.5 & 111	& n \\
86 & 10 7 9.01 & 15 43 27.3 & 19.62 & 18.86 & 1.2 & $-$21.5 & 21.5 & 115	& n \\
93 & 10 7 7.60 & 15 54 26.2 & 20.34 & 19.78 & 0.9 & $-$10.5 & 10.5 & $-$23	& n \\
95 & 10 7 7.65 & 16 02 43.2 & 18.53 & 17.78 & 0.9 & $-$2.2 & 2.4 & $-$41	& n \\
97 & 10 6 36.93 & 15 53 25.2 & 18.83 & 18.13 & $-$6.5 & $-$11.5 & 13.2 & 43	& n \\
98 & 10 7 14.58 & 16 01 54.5 & 18.84 & 18.07 & 2.5 & $-$3.0 & 3.9 & 199	& y \\
99 & 10 7 20.47 & 16 02 3.3 & 18.68 & 18.03 & 4.0 & $-$2.9 & 4.9 & 25	& n \\
101 & 10 6 59.01 & 15 44 18.9 & 19.75 & 19.13 & $-$1.2 & $-$20.6 & 20.6 & 307	& n \\
102 & 10 7 22.39 & 15 45 19.1 & 19.87 & 19.41 & 4.4 & $-$19.6 & 20.1 & 96	& n \\
104 & 10 6 44.62 & 15 47 1.9 & 17.44 & 16.50 & $-$4.7 & $-$17.9 & 18.5 & 41	& n \\
105 & 10 6 53.97 & 15 48 56.9 & 19.07 & 18.31 & $-$2.4 & $-$16.0 & 16.1 & $-$77	& n \\
107 & 10 7 6.88 & 15 44 42.6 & 19.45 & 18.62 & 0.7 & $-$20.2 & 20.2 & 154	& n \\
113 & 10 6 59.91 & 15 51 26.2 & 19.74 & 19.20 & $-$1.0 & $-$13.5 & 13.5 & $-$15	& n \\
117 & 10 5 57.65 & 15 41 51.9 & 17.52 & 16.64 & $-$16.0 & $-$23.0 & 28.0 & 173	& n \\
119 & 10 6 36.03 & 15 44 13.4 & 20.71 & 20.18 & $-$6.7 & $-$20.7 & 21.8 & 17	& n \\
123 & 10 6 6.83 & 15 45 36.6 & 19.11 & 18.43 & $-$13.8 & $-$19.3 & 23.7 & 114	& n \\
129 & 10 6 25.44 & 15 46 45.5 & 20.21 & 19.65 & $-$9.3 & $-$18.2 & 20.4 & 12	& n \\
130 & 10 5 49.10 & 15 37 43.9 & 21.49 & 21.05 & $-$18.0 & $-$27.2 & 32.6 & $-$68	& n \\
133 & 10 6 12.02 & 15 45 48.5 & 20.08 & 19.59 & $-$12.5 & $-$19.1 & 22.8 & 327	& n \\
137 & 10 6 26.43 & 15 49 37.6 & 20.11 & 19.43 & $-$9.0 & $-$15.3 & 17.8 & 82	& n \\
139 & 10 6 20.04 & 16 01 2.7 & 19.23 & 18.40 & $-$10.6 & $-$3.9 & 11.2 & 95	& n \\
141 & 10 5 28.24 & 15 34 40.7 & 19.91 & 19.44 & $-$23.1 & $-$30.2 & 38.0 & 57	& n \\
143 & 10 5 39.93 & 15 35 47.4 & 20.20 & 19.68 & $-$20.2 & $-$29.1 & 35.5 & $-$67	& n \\
147 & 10 6 6.39 & 15 47 9.5 & 19.38 & 18.72 & $-$13.9 & $-$17.8 & 22.5 & 35	& n \\
148 & 10 6 21.43 & 15 52 28.2 & 18.98 & 18.29 & $-$10.2 & $-$12.4 & 16.1 & 100	& n \\
154 & 10 5 3.32 & 15 40 48.3 & 17.52 & 16.60 & $-$29.0 & $-$24.1 & 37.7 & 95	& n \\
156 & 10 6 10.80 & 15 50 36.4 & 20.22 & 19.76 & $-$12.8 & $-$14.3 & 19.2 & $-$8	& n \\
157 & 10 6 36.70 & 15 54 13.0 & 20.27 & 19.76 & $-$6.6 & $-$10.7 & 12.6 & $-$20	& n \\
160 & 10 6 23.32 & 15 55 4.8 & 19.75 & 19.11 & $-$9.8 & $-$9.8 & 13.9 & 81	& n \\
161 & 10 6 39.33 & 16 00 8.6 & 19.46 & 18.91 & $-$5.9 & $-$4.8 & 7.6 & 269	& n \\
163 & 10 5 51.38 & 15 50 21.1 & 19.23 & 18.41 & $-$17.5 & $-$14.6 & 22.7 & 137	& n \\
173 & 10 5 30.59 & 15 54 18.1 & 19.80 & 19.29 & $-$22.5 & $-$10.6 & 24.8 & 305	& n \\
174 & 10 5 20.06 & 15 51 23.7 & 20.07 & 19.31 & $-$25.0 & $-$13.5 & 28.4 & 3	& n \\
175 & 10 5 0.99 & 15 52 59.0 & 19.62 & 19.07 & $-$29.6 & $-$11.9 & 31.9 & $-$15	& n \\
175 & 10 5 0.99 & 15 52 59.0 & 19.62 & 19.07 & $-$29.6 & $-$11.9 & 31.9 & $-$28	& n \\
177 & 10 5 27.03 & 15 56 34.2 & 18.39 & 17.66 & $-$23.3 & $-$8.3 & 24.8 & 94	& n \\
179 & 10 4 49.96 & 15 56 22.9 & 18.49 & 17.87 & $-$32.2 & $-$8.5 & 33.3 & 83	& n \\
180 & 10 6 21.88 & 15 53 3.4 & 19.18 & 18.51 & $-$10.1 & $-$11.9 & 15.6 & $-$23	& n \\
181 & 10 6 35.55 & 16 04 15.0 & 18.69 & 17.99 & $-$6.8 & $-$0.7 & 6.9 & 69	& n \\
182 & 10 6 10.62 & 15 58 30.7 & 17.40 & 16.53 & $-$12.8 & $-$6.4 & 14.3 & $-$11	& n \\
183 & 10 6 49.05 & 16 03 48.7 & 20.57 & 20.26 & $-$3.6 & $-$1.1 & 3.8 & 185	& y \\
188 & 10 6 34.93 & 15 57 16.2 & 17.32 & 16.46 & $-$7.0 & $-$7.6 & 10.4 & 72	& n \\
194 & 10 4 47.97 & 16 03 48.6 & 19.54 & 18.92 & $-$32.7 & $-$1.1 & 32.7 & 139	& n \\
195 & 10 4 28.53 & 15 55 59.1 & 18.05 & 17.31 & $-$37.4 & $-$8.9 & 38.4 & 76	& n \\
198 & 10 4 42.58 & 15 56 47.9 & 17.25 & 16.50 & $-$34.0 & $-$8.1 & 34.9 & 11	& n \\
201 & 10 5 20.71 & 16 06 36.7 & 18.03 & 17.07 & $-$24.8 & 1.7 & 24.9 & 95	& n \\
210 & 10 6 14.87 & 16 08 58.7 & 20.10 & 19.33 & $-$11.8 & 4.1 & 12.5 & 45	& n \\
211 & 10 6 47.73 & 16 04 25.2 & 17.98 & 17.03 & $-$3.9 & $-$0.5 & 3.9 & 34	& n \\
214 & 10 6 33.62 & 16 09 8.9 & 19.47 & 18.78 & $-$7.3 & 4.2 & 8.4 & 7	& n \\
217 & 10 6 0.16 & 16 05 18.6 & 20.20 & 19.51 & $-$15.3 & 0.4 & 15.3 & 242	& n \\
220 & 10 6 22.36 & 16 04 52.3 & 20.60 & 20.04 & $-$10.0 & $-$0.0 & 10.0 & 199	& y \\
222 & 10 5 17.23 & 16 20 35.6 & 18.08 & 17.43 & $-$25.6 & 15.7 & 30.0 & 105	& n \\
231 & 10 5 20.36 & 16 24 39.6 & 17.47 & 16.50 & $-$24.9 & 19.8 & 31.8 & $-$14	& n \\
237 & 10 6 57.78 & 16 07 35.8 & 18.73 & 17.96 & $-$1.5 & 2.7 & 3.1 & 86	& n \\
238 & 10 6 6.74 & 16 16 48.6 & 19.66 & 18.89 & $-$13.7 & 11.9 & 18.2 & 9	& n \\
240 & 10 6 18.00 & 16 12 20.9 & 20.14 & 19.45 & $-$11.0 & 7.4 & 13.3 & 146	& n \\
241 & 10 6 38.68 & 16 9 42.5 & 18.66 & 17.79 & $-$6.1 & 4.8 & 7.7 & 75	& n \\
242 & 10 6 8.75 & 16 12 52.2 & 18.57 & 17.91 & $-$13.3 & 8.0 & 15.5 & 69	& n \\
247 & 10 5 51.62 & 16 12 54.3 & 18.21 & 17.30 & $-$17.4 & 8.0 & 19.1 & 73	& n \\
251 & 10 6 29.98 & 16 20 12.2 & 18.14 & 17.45 & $-$8.2 & 15.3 & 17.3 & 35	& n \\
258 & 10 6 23.61 & 16 17 59.6 & 17.66 & 16.94 & $-$9.7 & 13.1 & 16.3 & 123	& n \\
259 & 10 6 13.78 & 16 16 58.6 & 19.62 & 18.84 & $-$12.1 & 12.1 & 17.1 & 21	& n \\
265 & 10 5 44.53 & 16 36 11.1 & 18.69 & 18.10 & $-$19.0 & 31.3 & 36.6 & $-$62	& n \\
270 & 10 6 20.86 & 16 26 24.9 & 19.74 & 19.07 & $-$10.3 & 21.5 & 23.9 & 226	& y \\
273 & 10 5 44.64 & 16 37 37.8 & 17.56 & 16.74 & $-$19.0 & 32.7 & 37.8 & 46	& n \\
274 & 10 6 28.54 & 16 16 33.6 & 17.78 & 16.93 & $-$8.5 & 11.6 & 14.4 & $-$44	& n \\
275 & 10 5 40.52 & 16 38 55.1 & 17.97 & 17.21 & $-$20.0 & 34.0 & 39.5 & 71	& n \\
276 & 10 6 28.86 & 16 35 6.3 & 18.74 & 18.13 & $-$8.4 & 30.2 & 31.3 & 91	& n \\
278 & 10 6 26.25 & 16 29 45.9 & 18.34 & 17.70 & $-$9.0 & 24.9 & 26.4 & 46	& n \\
280 & 10 6 11.70 & 16 24 58.6 & 18.87 & 18.12 & $-$12.5 & 20.1 & 23.7 & $-$37	& n \\
295 & 10 7 7.58 & 16 38 49.6 & 17.89 & 17.03 & 0.9 & 33.9 & 33.9 & $-$9	& n \\
299 & 10 6 58.49 & 16 20 45.6 & 19.84 & 19.20 & $-$1.3 & 15.8 & 15.9 & 295	& n \\
302 & 10 6 38.05 & 16 42 45.2 & 17.73 & 16.77 & $-$6.2 & 37.8 & 38.3 & $-$46	& n \\
311 & 10 7 7.57 & 16 43 48.2 & 17.76 & 16.81 & 0.9 & 38.9 & 38.9 & 10	& n \\
312 & 10 7 39.91 & 16 29 21.3 & 18.71 & 17.84 & 8.6 & 24.4 & 25.9 & 141	& n \\
318 & 10 7 10.08 & 16 06 23.9 & 19.19 & 18.43 & 1.5 & 1.5 & 2.1 & 213	& y \\
325 & 10 7 29.83 & 16 29 50.7 & 17.59 & 16.60 & 6.2 & 24.9 & 25.7 & 35	& n \\
326 & 10 7 39.36 & 16 18 19.4 & 18.05 & 17.30 & 8.5 & 13.4 & 15.9 & 105	& n \\
327 & 10 8 20.61 & 16 34 51.7 & 18.26 & 17.48 & 18.4 & 30.0 & 35.1 & $-$5	& n \\
331 & 10 7 43.47 & 16 22 59.3 & 18.28 & 17.41 & 9.5 & 18.1 & 20.4 & 42	& n \\
333 & 10 8 19.15 & 16 23 46.8 & 19.08 & 18.45 & 18.0 & 18.9 & 26.1 & 16	& n \\
335 & 10 7 30.00 & 16 20 11.8 & 17.37 & 16.56 & 6.2 & 15.3 & 16.5 & 64	& n \\
338 & 10 8 39.66 & 16 28 26.7 & 19.16 & 18.58 & 22.9 & 23.6 & 32.9 & 288	& n \\
340 & 10 8 11.98 & 16 38 15.2 & 17.82 & 16.99 & 16.3 & 33.3 & 37.1 & 4	& n \\
342 & 10 8 1.37 & 16 22 17.3 & 18.54 & 17.82 & 13.8 & 17.4 & 22.2 & 37	& n \\
343 & 10 8 33.19 & 16 37 52.3 & 19.10 & 18.50 & 21.4 & 33.0 & 39.3 & 144	& n \\
345 & 10 8 46.20 & 16 29 27.8 & 17.69 & 16.83 & 24.5 & 24.6 & 34.7 & $-$20	& n \\
346 & 10 8 11.13 & 16 23 9.1 & 19.50 & 18.94 & 16.1 & 18.2 & 24.3 & 98	& n \\
352 & 10 8 16.68 & 16 31 38.9 & 17.76 & 17.00 & 17.4 & 26.7 & 31.9 & 102	& n \\
355 & 10 8 25.17 & 16 16 51.2 & 19.18 & 18.48 & 19.5 & 12.0 & 22.9 & 24	& n \\
356 & 10 7 22.63 & 16 08 47.3 & 18.83 & 18.10 & 4.5 & 3.9 & 5.9 & $-$12	& n \\
358 & 10 9 7.00 & 16 27 52.9 & 19.39 & 18.76 & 29.5 & 23.0 & 37.4 & 133	& n \\
358 & 10 9 7.00 & 16 27 52.9 & 19.39 & 18.76 & 29.5 & 23.0 & 37.4 & 145	& n \\
370 & 10 8 20.27 & 16 17 32.4 & 17.70 & 16.75 & 18.3 & 12.6 & 22.2 & $-$23	& n \\
371 & 10 8 10.23 & 16 11 32.0 & 18.41 & 17.75 & 15.9 & 6.6 & 17.2 & 22	& n \\
372 & 10 8 8.44 & 16 21 54.5 & 19.07 & 18.27 & 15.5 & 17.0 & 23.0 & 102	& n \\
382 & 10 7 47.17 & 16 08 37.8 & 18.42 & 17.62 & 10.4 & 3.7 & 11.0 & $-$60	& n \\
387 & 10 8 28.30 & 16 09 12.8 & 19.34 & 18.74 & 20.2 & 4.3 & 20.7 & 14	& n \\
388 & 10 8 22.51 & 16 05 14.2 & 17.97 & 17.23 & 18.9 & 0.3 & 18.9 & $-$14	& n \\
388 & 10 8 22.51 & 16 05 14.2 & 17.97 & 17.23 & 18.9 & 0.3 & 18.9 & $-$7	& n \\
390 & 10 8 52.27 & 16 07 45.7 & 18.22 & 17.37 & 26.0 & 2.9 & 26.2 & 35	& n \\
391 & 10 8 50.08 & 16 01 52.3 & 17.89 & 17.03 & 25.5 & $-$3.0 & 25.7 & $-$13	& n \\

\enddata
\end{deluxetable}

\clearpage
\begin{deluxetable}{rrrccccccccr}
\tablecaption{OBSERVATIONAL DATA AND DERIVED ATMOSPHERIC PARAMETERS FOR CANDIDATE MEMBER STARS OF BO\"{O}TES I\vspace{0mm}}
\tablewidth{0pt}
\tablehead{
\colhead {Star} & {R}           & {$v_{helio}$}      & {g}  & {(g-r)$_{0}$}  & {(r-z)$_{0}$}  & {K$^{\prime}$} & {\teff} & {\logg}  & {[Fe/H]} & {[C/H]} & {[C/Fe]}   \\
         {}     & {($\arcmin$)} & {(kms$^{-1}$)} & {}         & {}             & {}       & {({\AA})}       & {(K)}           & {(cgs)}    & {}       & {}       & {}        \\ 
         {(1)}  & {(2)}         & {(3)}          & {(4)}      & {(5)}          & {(6)}    & {(7)}           & {(8)}           & {(9)}      & {(10)}   & {(11)}   & {(12)}      
}
\startdata
7    & 12.1 &  106 & 18.30 & 0.693 & 0.450 & 7.49 & 4800 & 1.6 & $-$2.32\tablenotemark{a} & --2.82 & --0.50 \\  
8    & 12.6 &  106 & 19.02 & 0.551 & 0.349 & 4.53 & 5090 & 2.3 & $-$2.75\tablenotemark{a} & --2.15 &   0.60 \\   
9    & 11.1 &  112 & 17.92 & 0.803 & 0.538 & 6.62 & 4630 & 1.1 & $-$2.67\tablenotemark{a} & --3.22 & --0.55 \\   
33   &  5.2 &  107 & 18.23 & 0.736 & 0.474 & 5.17 & 4730 & 1.4 & $-$2.29\tablenotemark{b} & --1.99 &   0.30 \\  
34   & 13.1 &  111 & 18.74 & 0.647 & 0.452 & 4.94 & 4840 & 1.6 & $-$3.10\tablenotemark{a} & --2.55 &   0.55 \\   
41   &  6.2 &  105 & 18.38 & 0.697 & 0.475 & 8.13 & 4750 & 1.6 & $-$1.93\tablenotemark{b} & --2.58 & --0.65 \\   
78   & 16.9 &  105 & 19.30 & 0.619 & 0.391 & 6.10 & 4950 & 1.9 & $-$2.46\tablenotemark{a} & --2.61 & --0.15 \\   
94   &  7.4 &   94 & 17.52 & 0.872 & 0.575 & 6.63 & 4570 & 0.8 & $-$2.90\tablenotemark{b} & --3.35 & --0.45 \\    
117  &  2.1 &   96 & 18.21 & 0.746 & 0.492 & 8.78 & 4700 & 1.4 & $-$2.25\tablenotemark{b} & --2.55 & --0.30 \\  
121  & 12.0 &  112 & 17.92 & 0.811 & 0.525 & 7.62 & 4630 & 1.1 & $-$2.37\tablenotemark{a} & --2.62 & --0.25 \\   
127  &  6.2 &   99 & 18.16 & 0.773 & 0.493 & 9.47 & 4670 & 1.4 & $-$2.08\tablenotemark{b} & --2.33 & --0.25 \\ 
130  &  4.1 &  111 & 18.21 & 0.707 & 0.475 & 6.60 & 4750 & 1.4 & $-$2.29\tablenotemark{b} & --2.69 & --0.40 \\ 
911  &  7.0 &   94 & 17.96 & 0.793 & 0.597 & 8.47 & 4540 & 1.1 & $-$1.98\tablenotemark{a} & --2.53 & --0.55 \\  
980  & 48.8 &  106 & 18.49 & 0.610 & 0.437 & 4.14 & 4890 & 1.7 & $-$3.09\tablenotemark{a} & --3.09 &   0.00 \\   
1069 & 29.0 &  113 & 19.05 & 0.553 & 0.371 & 5.57 & 5050 & 2.2 & $-$2.51\tablenotemark{a} & --2.86 & --0.35 \\  
1137 & 24.0 &  108 & 18.10 & 0.718 & 0.517 & 3.06 & 4710 & 1.2 & $-$3.66\tablenotemark{c} & --3.46 &   0.20 \\    
\enddata
\tablenotetext{a}{From this work, based on AAOmega moderate-resolution Ca II K$^{\prime}$ data}
\tablenotetext{b}{From Gilmore et al.\ (2010, in prep.), based on VLT/UVES/Flames data}
\tablenotetext{c}{From Norris et al. \ (2010)}
\end{deluxetable}

\clearpage
\begin{deluxetable}{rrcccccccrrr}
\tablecaption{OBSERVATIONAL DATA AND DERIVED ATMOSPHERIC PARAMETERS FOR CANDIDATE MEMBER STARS OF SEGUE~1\vspace{0mm}}
\tablewidth{0pt}
\tablehead{
\colhead {Star} & {R}           & {$v_{helio}$}      & {g}   & {(g-r)$_{0}$}  & {(r-z)$_{0}$}   & {K$^{\prime}$} & {\teff} & {\logg}  & {[Fe/H]} & {[C/H]} & {[C/Fe]}   \\
         {}     & {($\arcmin$)} & {(kms$^{-1}$)} & {}    & {}             & {}              & {({\AA})}      & {(K)}           &  {(cgs)}   & {}       & {}       & {}        \\ 
         {(1)}  & {(2)}         & {(3)}          & {(4)} & {(5)}          & {(6)}           & {(7)}          & {(8)}           & {(9)}      & {(10)}    & {(11)}   & {(12)}      
}
\startdata
7                     & 16.9 &  194  & 18.05 & 0.596 & 0.354 & 1.86  & 4960 & 1.9 & $-3.5$\tablenotemark{a}   & $-$1.2         &  $+2.3$ \\  
31                    & 10.0 &  211  & 18.61 & 0.436 & 0.293 & 6.00  & 5420 & 3.3 & $-1.9$\tablenotemark{b}   & $-$1.9         &  $0.0$  \\    
42\tablenotemark{c}   &  9.4 &  195  & 18.75 & 0.441 & 0.204 & 7.71  & 5570 & 3.7 & $-1.5$\tablenotemark{b}   &   ...          &     ... \\ 
71                    & 14.0 &  212  & 18.49 & 0.527 & 0.282 & 6.77  & 5200 & 2.6 & $-2.2$\tablenotemark{b}   & $-$2.4         &  $-0.2$ \\  
98\tablenotemark{d}   &  3.9 &  199  & 18.86 & 0.528 & 0.284 & 0.65  & 5150 & 2.4 & ...                       & ...            & ...     \\\\
3451364\tablenotemark{e} &  3.6 &  216\tablenotemark{e} & 18.9\tablenotemark{e}  & 0.48\tablenotemark{e}  & ...  & ...   & 5190\tablenotemark{e}  & 2.8\tablenotemark{e} & $-3.3$\tablenotemark{e}  & ...            & ...    \\
\\
 
\enddata
\tablenotetext{a}{From the new VLT/UVES high-resolution, high-$S/N$, data mentioned in the text. The abundance coincides with the value obtained from AAOmega moderate-resolution Ca II K$^{\prime}$ data.}
\tablenotetext{a}{From this work, based on AAOmega Ca II K$^{\prime}$ data.}
\tablenotetext{c} {\ Likely non-member, though velocity matches systemic.  See text for discussion.}
\tablenotetext{d}{Inconsistent blue and red abundances.  See text for discussion.}
\tablenotetext{e}{RGB member from Geha et al.\ (2009), not observed by us.}
\end{deluxetable}

\clearpage
\clearpage
\begin{deluxetable}{lccccccrc}
\tablecaption{ABSOLUTE MAGNITUDES, MEAN METALLICITIES, \&  METALLICITY DISPERSIONS FOR DWARF SPHEROIDAL GALAXIES AND $\omega$~CENTAURI}
\tablewidth{0pt}
\tablehead{
\colhead {Galaxy} & {M$_{V,\,total}$} & {Source}\tablenotemark{a}         & {$\langle$[Fe/H]$\rangle$}  & {s.e.($\langle$[Fe/H]$\rangle$)}  & {$\sigma$([Fe/H])}      & {s.e.($\sigma$([Fe/H)])}    &  {No.}   & {Source}\tablenotemark{a}  \\
         {(1)}  &     {(2)}              &       {(3)}                &         {(4)}                    &           {(5)}        &           {(6)}            &   {(7)}           &   {(8)}  &   {(9)}   
}
\startdata
Fornax   &    --13.2 & 1 &   --0.88 &    0.06 &    0.35 &    0.04 &     36  &     2 \\
Sculptor &    --11.1 & 1 &   --1.56 &    0.04 &    0.37 &    0.03 &     93  &     2 \\
{\wcen}  &    --10.3 & 3 &   --1.60 &    0.01 &    0.27 &    0.01 &    517  &     4 \\ 
Sextans  &     --9.5 & 1 &   --2.08 &    0.03 &    0.37 &    0.02 &    202  &     5 \\
Carina   &     --9.3 & 1 &   --1.81 &    0.02 &    0.29 &    0.01 &    364  &     5 \\
UMi      &     --8.9 & 1 &   --2.23 &    0.04 &    0.31 &    0.03 &     70  &     6 \\
Draco    &     --8.8 & 1 &   --2.19 &    0.04 &    0.32 &    0.02 &     95  &     6 \\
CVn I    &     --8.6 & 7 &   --2.08 &    0.04 &    0.41 &    0.02 &    165  &     8 \\
Her      &     --6.6 & 7 &   --2.58 &    0.11 &    0.47 &    0.08 &     20  &     8 \\
Bootes I &     --6.3 & 7 &   --2.55 &    0.11 &    0.37 &    0.08 &     16  &     9 \\  
UMa I    &     --5.5 & 7 &   --2.29 &    0.10 &    0.50 &    0.07 &     28  &     8 \\
Leo IV   &     --5.0 & 7 &   --2.58 &    0.22 &    0.72 &    0.15 &     12  &     8 \\
CVn II   &     --4.9 & 7 &   --2.19 &    0.14 &    0.54 &    0.10 &     16  &     8 \\
UMa II   &     --4.2 & 7 &   --2.44 &    0.16 &    0.53 &    0.11 &     13  &     8 \\
Com      &     --4.1 & 7 &   --2.53 &    0.09 &    0.40 &    0.06 &     24  &     8 \\
Seg 1    &     --1.5 & 7 &   --2.72 &    0.40 &    0.70 &    0.29 &      4  &     9,10 \\ 
\enddata

\tablenotetext{a}{1. Mateo 1998; 2. Battaglia et al. (2008);
  3. Harris,
  W.E. \texttt{http://physwww.physics.mcmaster.ca/~harris/mwgc.dat};
  4. Norris, Freeman \& Mighell 1996 (assuming [Ca/Fe] = +0.4 for
  [Fe/H] $<$ --1.0, and linearly decreasing [Ca/Fe] from +0.4 to 0.0
  between [Fe/H] = --1.0 and 0.0); 5. Helmi et al.\ (2006); 6. Winnick
  (2003); 7. Martin et al.\ (2008); 8. Kirby et al.\ (2008); 9. This
  work; 10. Geha et al.\ (2009)}

\end{deluxetable}

\clearpage

\clearpage

\clearpage

\clearpage

\clearpage 

\clearpage 

\clearpage 

\clearpage 

\clearpage 

\begin{figure}[htbp]
\vspace{1cm}
\begin{center}
\includegraphics[width=7cm,angle=0]{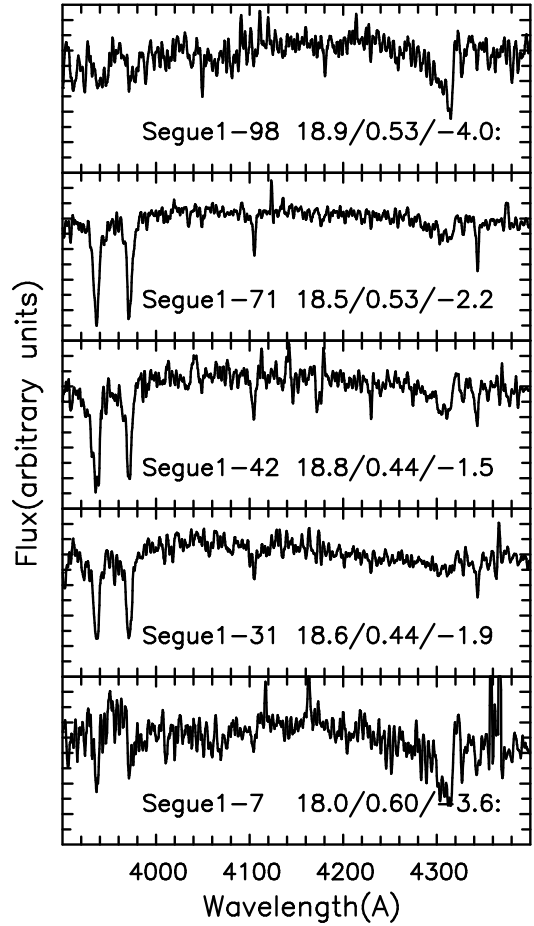}

\caption{\label{Fig:Segue1_Spectra} Spectra of the five candidate radial
velocity members of {\seg} (continuum-normalized and broadened to a
resolution of FWHM = 2.5\ {\AA}).  Panels are labeled with star name and
$g$/{\grz}/[Fe/H].}

\end{center}
\end{figure}

\clearpage

\begin{figure}[htbp]
\vspace{1cm}
\begin{center}
\includegraphics[width=7cm,angle=0]{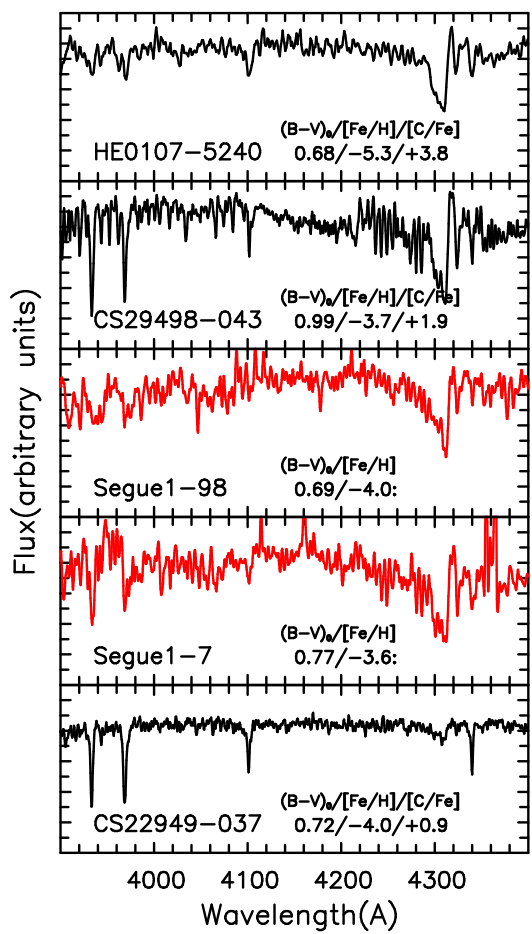}

\caption{\label{Fig:Segue1_UMP} Comparison of the spectra of the
putative C-rich stars {\segs}--7 and {\segs}--98 with those of
extremely Fe-poor, C-rich Galactic halo giants of similar color
({\teff}) (from ANU's 2.3m telescope).  The spectra have been
continuum-normalized and broadened to resolution of FWHM = 2.5\ {\AA}.
Individual panels list (B-V)$_{0}$/[Fe/H]/[C/H], as indicated, from
Aoki et al. (2002), Christlieb et al. (2004), McWilliam et al.\
(1995), and the present work.}

\end{center}
\end{figure}

\clearpage

\begin{figure}[htbp]
\vspace{1cm}
\begin{center}
\includegraphics[width=7cm,angle=0]{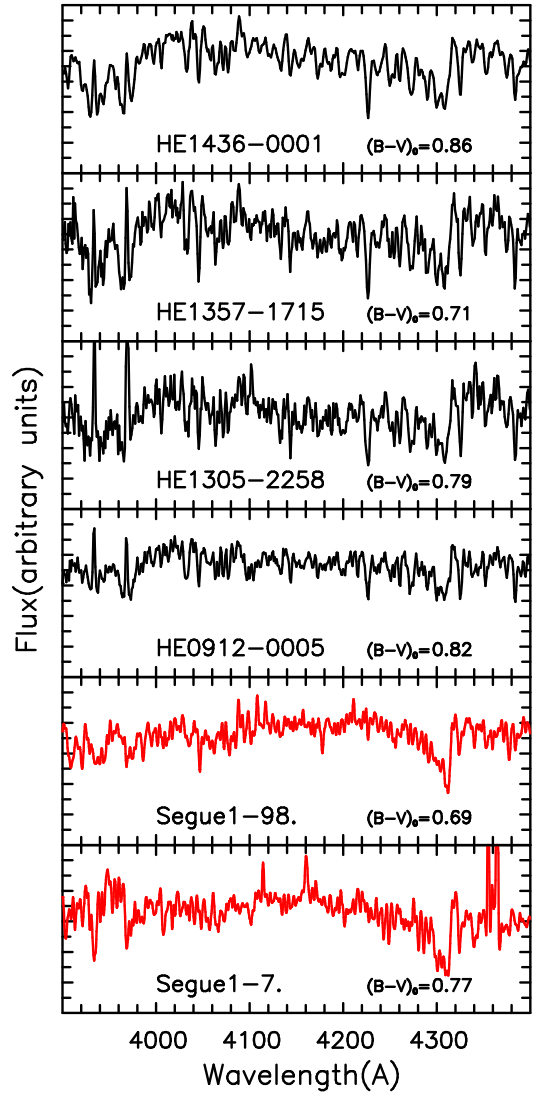}

\caption{\label{Fig:Segue1_HKem} Comparison of the spectra of
  {\segs}--7 and {\segs}--98 with those of extremely metal-poor stars from
  the Hamburg ESO Survey found serendipitously to have Ca~II H\&K
  emission cores.  See text for discussion.}

\end{center}
\end{figure}

\clearpage

\begin{figure}[htbp]
\vspace{1cm}
\begin{center}
\includegraphics[width=7cm,angle=0]{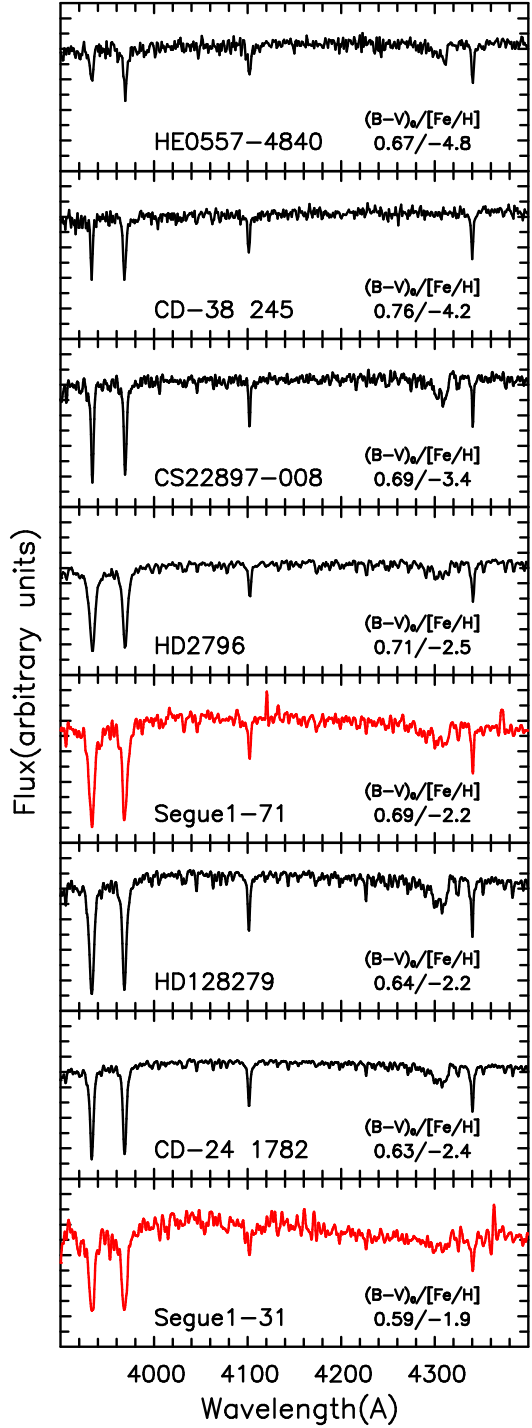}

\caption{\label{Fig:Segue1_normal} Comparison of the spectra of the
  C-normal {\seg} stars with C-normal Galactic halo giants of similar
  color ({\teff}) (from ANU's 2.3m telescope).  The spectra have been
  continuum-normalized and broadened to resolution of FWHM = 2.5\ {\AA}.
  Individual panels list (B-V)$_{0}$/[Fe/H] from Cayrel et al.\
  (2004), Chiba \& Yoshii (1998), and Norris et al.\ (1985), as
  described in the text.}

\end{center}
\end{figure}

\clearpage

\begin{figure}[htbp]
\vspace{1cm}
\begin{center}
\includegraphics[width=10cm,angle=270]{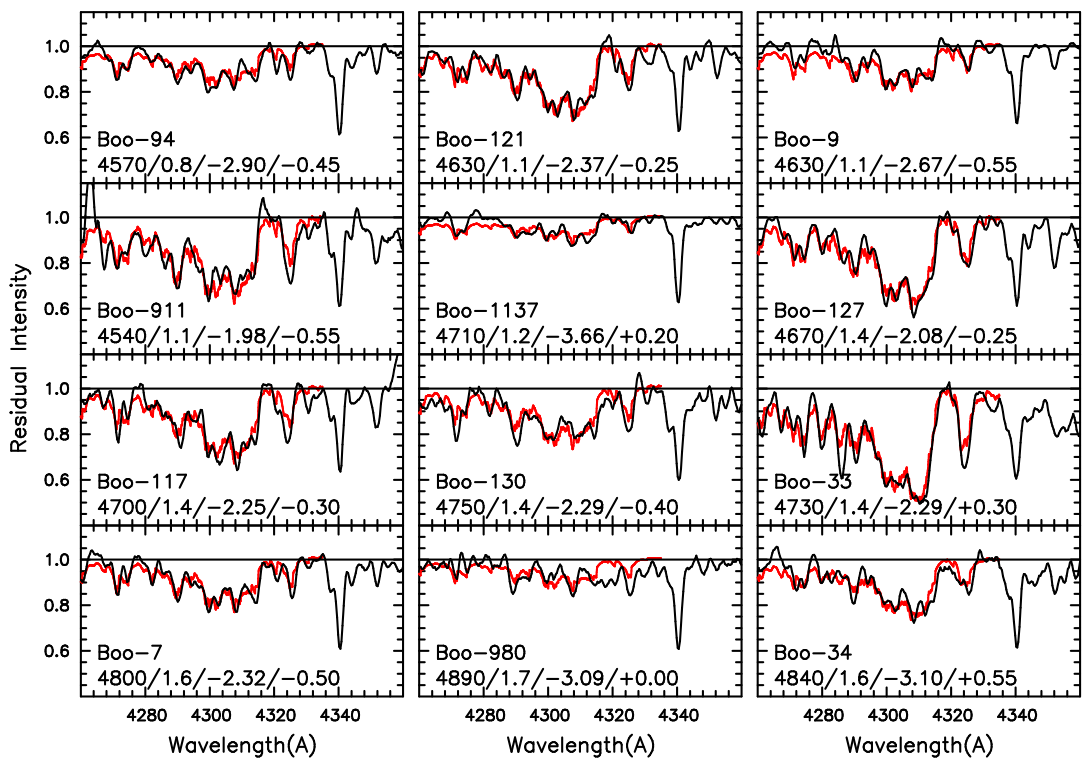}

\caption{\label{Fig:Bootes_Carbon} Comparison of the observed spectra
(black lines) of 12 red giants in {\boo} with the best fitting
synthetic spectra (red lines) in the region of the G-band of the CH
molecule. The stellar identification and T$_{\rm
eff}$/log~$g$/[Fe/H]/[C/Fe] are presented in each panel. (The
corresponding spectra of the stars in the range 3900--4400\ {\AA} may
be found in Figure~1 of Norris et al.\ (2008).)}

\end{center}
\end{figure}

\clearpage
\begin{figure}[htbp]
\vspace{1cm}
\begin{center}
\includegraphics[width=7cm,angle=0]{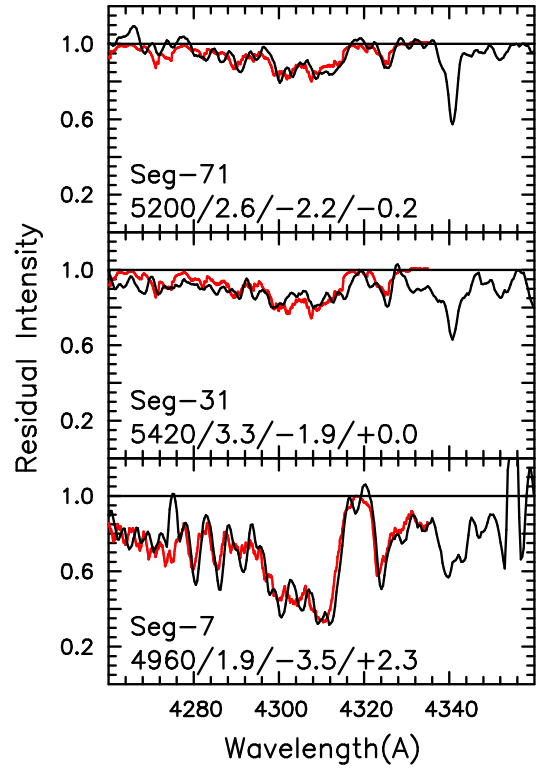}

\caption{\label{Fig:Segue_Carbon} Comparison of the observed spectra
(black lines) of the three putative radial velocity members of {\seg}
with the best fitting synthetic spectra (red lines) in the region of
the G-band. Stellar identification and T$_{\rm
eff}$/log~$g$/[Fe/H]/[C/Fe] are presented in each panel.}

\end{center}
\end{figure}

\clearpage

\begin{figure}[htbp]
\begin{center}
\includegraphics[width=7cm,angle=0]{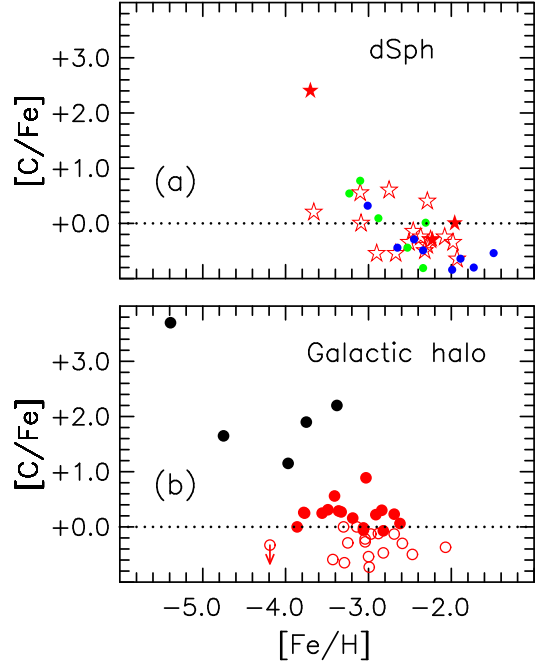}

\caption{\label{Fig:[C/Fe]_[Fe/H]} [C/Fe] as a function of [Fe/H] for
Galactic dwarf satellite (upper panel) and halo (lower panel) red
giants.  In the upper panel the {\uf} {\boo} and {\seg} are
represented by open and filled red stars, respectively, while the
filled green and blue circles come from Frebel et al.\ (2010a) for the
{\uf} UMa II and Com, and from Cohen \& Huang (2009) for the Draco
dSph, respectively.  The open and filled red circles in the lower
panel represent the ``mixed'' and ``unmixed'' stars of Spite et al.\
(2005), while the filled black symbols represent the C-rich extremely
iron-poor giants HE0107--5240 ([Fe/H] = --5.40, Christlieb et al\
2004), HE0557--4840 ([Fe/H] = --4.75, Norris et al.\ 2007),
CS22949--037 ([Fe/H] = --3.97, Spite et al.\ 2005), CS29498--043
([Fe/H] = --3.75, Aoki et al.\ 2002) and CS22957--027 ([Fe/H] =
--3.38, Norris et al.\ 1997).}

\end{center}
\end{figure}

\clearpage

\begin{figure}[htbp]
\begin{center}
\includegraphics[width=6cm,angle=0]{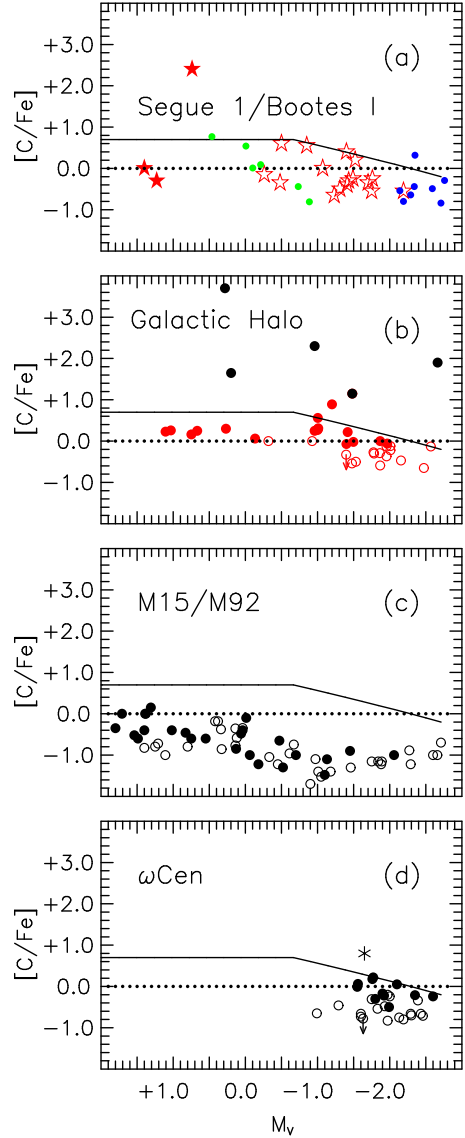}

\caption{\label{Fig:[C/Fe]_Mv} The dependence of [C/Fe] on absolute
magnitude, M$_{V}$, for (a) the {\uf} {\boo} and {\seg}
(open and filled stars), respectively, and UMa II and Com (filled
green circles) from Frebel et al.\ (2010a), together with data from
Cohen \& Huang (2009) for the Draco dSph (filled blue circles), (b) extremely
metal-poor giants of the Galactic halo (symbols as in
Figure~\ref{Fig:Segue_Carbon}), (c) the globular clusters M15 (open
symbols) and M92 (filled symbols; from Langer et al.\ 1986 and
Trefzger et al.\ 1983), and (d) {\wcen}, where filled and
open circles refer to CO-strong and CO-weak stars, respectively, and
the asterisk stands for a CH star (from Norris \& Da Costa 1995). The
horizontal dotted line lies at the solar value, [C/Fe] = 0.0, while
the continuous line corresponds to the division of Aoki et al.\ (2007)
between C-rich and C-normal stars.  See text for discussion.}

\end{center}
\end{figure}

\clearpage

\begin{figure}[htbp]
\begin{center}
\includegraphics[width=12cm,angle=0]{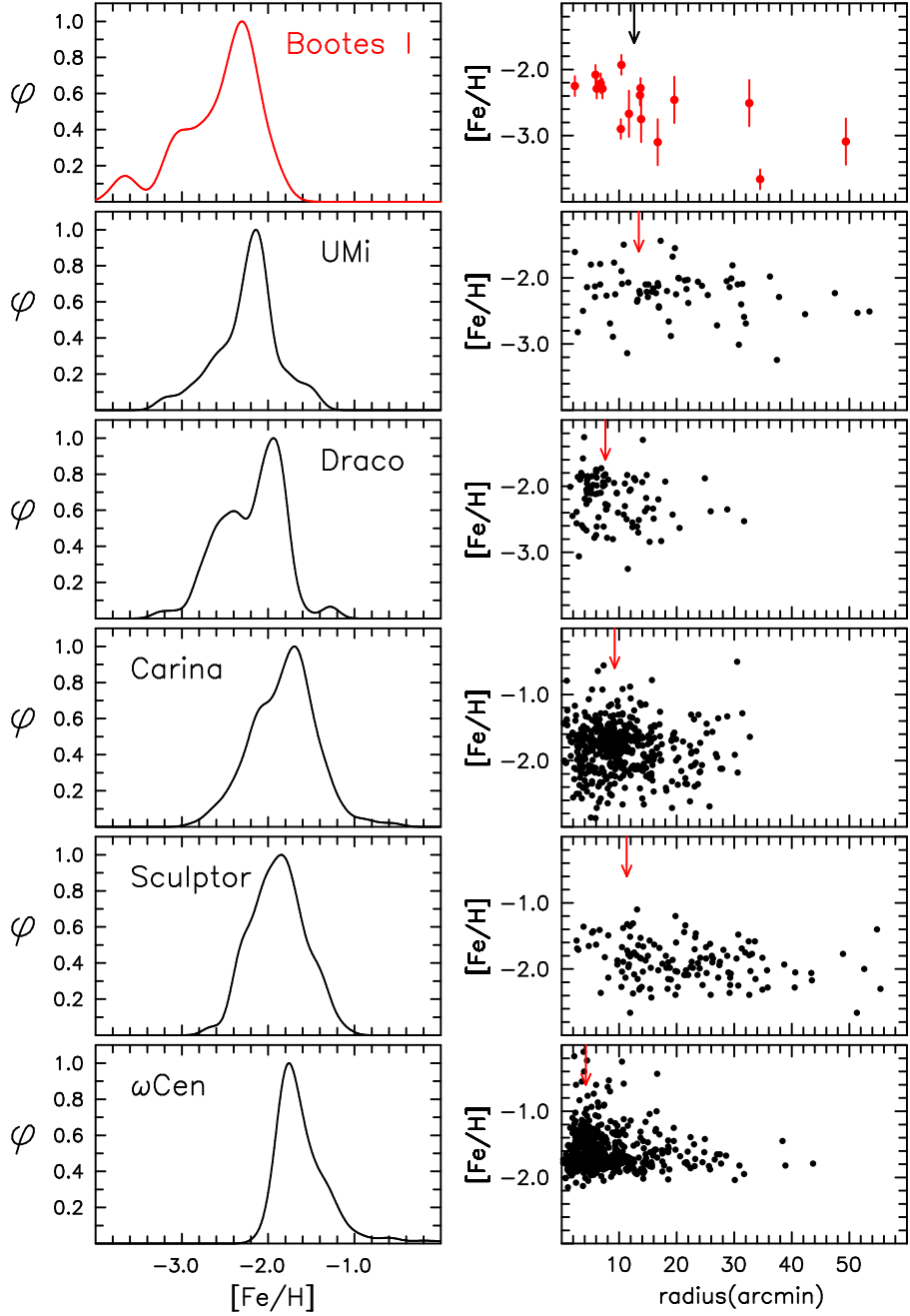}

\caption{\label{Fig:MDF} Left: Metallicity distribution functions for
{\boo} and Galactic dwarf spheroidal satellites. Right: [Fe/H] vs
elliptical radial distance for {\boo} and the other systems shown in
the left panels (except for {\wcen}, for which radial distance is
plotted). The arrows on the upper x-axes indicate the position of half-light
radius. For the other systems, the data have been taken from Winnick
(2003, UMi and Draco), Koch et al.\ (2006, Carina), Battaglia et al.\
(2008, Sculptor), and Norris et al.\ (1996, {\wcen}).}

\end{center}
\end{figure}

\clearpage

\begin{figure}[htbp]
\vspace{1cm}
\begin{center}
\includegraphics[width=7cm,angle=0]{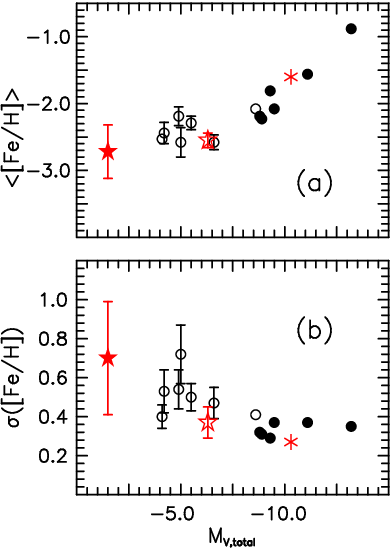}

\caption{\label{Fig:SigFe_vs_Mv} (a) Mean metallicity,
$\langle$[Fe/H]$\rangle$, and (b) metallicity dispersion,
$\sigma$[Fe/H], versus system absolute magnitude, M$_{V,\,total}$, for
the dwarf galaxy satellites of the Milky Way and {\wcen}.  The filled
and open circles refer to values for the dSphs and {\uf} dwarf
galaxies from the literature, respectively while the open and filled
stars represent {\boo} and {\seg}, respectively, from the present
work. The parameters for {\seg} assume all four stars discussed in the
text ({\segs}--7, 31, 71, and 3451364) are indeed members. {\wcen} is
shown as the asterisk.  See Table~5 for data and their sources.  Error
bars are not plotted when they are smaller than 0.10 and 0.05 for
$\langle$[Fe/H]$\rangle$ and $\sigma$[Fe/H], respectively.}

\end{center}
\end{figure}


\newpage

\begin{references}

\reference {} Abazajian, K.N. et al.\ 2009, ApJS, 182, 543 

\reference {}An, D. et al.\ 2008, ApJS, 179, 326

\reference {} Aoki, W., Beers, T.C., Christlieb, N., Norris, J.E.,
Ryan, S.G., \& Tsangarides, S. 2007, ApJ, 655, 492

\reference {} Aoki, W., Norris, J.E., Ryan, S.G., Beers, T.C., \&
Ando, H. 2002, ApJ, 576, L141

\reference {} Barry, D.C. 1988, ApJ, 334, 436

\reference {} Battaglia, G., Irwin, M., Tolstoy, E., Hill, V., Helmi,
A., Letarte, B., \& Jablonka, P. 2008, MNRAS, 383, 183

\reference {} Beers, T.C. \& Christlieb, N. 2005, ARA\&A,  43, 531

\reference {} Beers, T.C., Preston, G.W., \& Shectman, S.A. 1992, AJ,
103, 1987

\reference {} Beers, T.C., Rossi, S., Norris, J.E., Ryan, S.G., \&
Shefler, T.\ 1999, AJ, 117, 981

\reference {} Bekki, K. \& Freeman, K.C. 2003, MNRAS, 346, L11

\reference {} Bekki, K. \& Norris, J.E. 2006, ApJ, 637, L109

\reference {} Belokurov, V. et al.\ 2006, ApJ, 647, L111

\reference {} Belokurov, V. et al.\ 2007, ApJ, 654, 897

\reference {} Carollo, D. et al.\ 2007, Nature, 450, 1020

\reference {} Castelli, F. \& Kurucz, R.L. 2003, in IAU Symposium 210
``Modelling of Stellar Atmospheres'', eds. N. Piskunov, W.W. Weiss, \&
D.F. Gray (San Francisco: ASP) p.A20 (astro-ph/0405087)

\reference {} Cayrel, R. et al.\ 2004, A\&A, 416, 1117

\reference {} Chiba, M. \& Yoshii, Y. 1998, AJ, 115, 168

\reference {} Chou, M.-Y. et al.\ 2007, ApJ, 670, 346

\reference {} Christlieb, N., Bessell, M.S., Beers, T.C., Gustafsson,
B., Korn, A., Barklem, P.S., Karlsson, T., Mizuno-Wiedner, M., \&
Rossi, S. 2002, Nature, 419, 904

\reference {} Christlieb, N., Gustafsson, B., Korn, A.J., Barklem,
P.S., Beers, T.C., Bessell, M.S., Karlsson, T., \& Mizuno-Wiedner,
M. 2004, ApJ, 603, 708

\reference {} Cohen, J.G. \& Huang, W. 2009, ApJ, 701,1053

\reference {} Cottrell, P.L. \& Norris, J. 1978, ApJ, 221, 893

\reference {} Da Costa, G.S., Freeman, K.C., Kalnajs, A.J., Rodgers,
A.W., \& Stapinski, T.E. 1977, AJ, 82, 810

\reference {} Demarque, P., Woo, J.-H., Kim, Y.-C., \& Yi, S.K. 2004,
ApJS, 155, 667

\reference {} Feltzing, S., Eriksson, K., Kleyna, J., \&
Wilkinson, M.I. 2009, A\&A, 508, L1

\reference {} Frebel, A. et al.\ 2005, Nature, 434, 871

\reference {} Frebel, A., Kirby, E.N., \& Simon, J.D. 2010b, Nature, 464, 72

\reference {} Frebel, A., Simon, J.D., Geha, M., \& Willman, B. 2010a,
ApJ, 708, 560

\reference {} Fulbright, J.P., Rich, R.M., \& Castro, S. 2004, ApJ, 612,
447

\reference {} Geha, M., Willman, B., Simon, J.D., Strigari, L.E.,
Kirby, E.N., Law, D.R., \& Strader, J. 2009, ApJ, 692, 1464

\reference {} Gnedin, O.Y., Zhao, H., Pringle, J.E., Fall, S.M.,
Livio, M., \& Meylan, G. 2002, ApJ, 568, L23

\reference {} Gratton, R.G., Sneden, C., Carretta, E., \& Bragaglia,
A. 2000, A\&A, 354, 169

\reference {} Heavens, A.F. 1993, MNRAS, 263, 735

\reference {} Helmi, A. et al.\ 2006, ApJ, 651, L121

\reference {} Iwamoto, N., Umeda, H., Tominaga, N., Nomoto, K., \&
Maeda, K. 2005, Science, 309, 451

\reference {} Joggerst, C.C., Almgren, A., Bell, J., Heger, A.,
Whalen, D. \& Woosley, S.E. 2010, ApJ, 709, 11

\reference {} Kirby, E.N., Simon, J.D., Geha, M., Guhathakurta, P.,
\& Frebel, A. 2008, ApJ, 685, L43

\reference {} Klypin, A., Kravtsov, A.V., Valenzuela, O., \&
Prada, F. 1999, ApJ, 522, 82
 
\reference {} Koch, A., Grebel, E.K., Wyse, R.F.G., Kleyna, J.T.,
Wilkinson, M.I., Harbeck, D.R., Gilmore, G.F., \& Evans, N.W, 2006,
AJ, 131, 895

\reference {} Koch, A., McWilliam, A., Grebel, E.K., Zucker, D.B., \&
Belokurov, V. 2008, ApJ, 688, L13

\reference {} Kraft, R.P. \& Ivans, I.I. 2003, PASP, 115, 143

\reference {} Kurucz, R.L. 1993, CD-ROM 13, ATLAS9 Stellar Atmospheres
Programs and 2 km/s Grid (Cambridge: SAO)

\reference {} Langer, G.E., Kraft, R.P., Carbon, D.F., Friel, E., \&
Oke, J.B. 1986, PASP, 98, 473

\reference {} Marcolini, A., Sollima, A., D'Ercole, A., Gibson, B.K.,
\& Ferraro, F.R. 2007, MNRAS, 382, 443

\reference {} Martin, N.F., Ibata, R.F., Chapman, S.C., Irwin, M., \&
Lewis, G.F. 2007, MNRAS, 380, 281

\reference {} Martin, N.F., de Jong, J.T.A., \& Rix, H.-W. 2008,
ApJ, 684, 1075

\reference {} Mateo, M. 1998, ARA\&A, 36, 435

\reference {} McWilliam, A., Preston, G.W., Sneden, C., \& Searle,
L. 1995, AJ, 109, 2757

\reference {} Moore, B., Ghigna, S., Governato, F., Lake, G., Quinn,
T., Stadel, J., \& Tozzi, P. 1999, ApJ, 524, L19

\reference {} Mu\~{n}oz, R.R., Carlin, J.L., Frinchaboy, P.M.,
Nidever, D.L., Majewski, S.R., \& Patterson, R.L. 2006, ApJ, 650, L51

\reference {} Niederste-Ostholt, M., Belokurov, V., Evans, N.W.,
Gilmore, G., Wyse, R.F.G., \& Norris, J.E. 2009, MNRAS, 398, 1771

\reference {} Norris, J., Bessell, M.S., \& Pickles, A.J. 1985, ApJS,
58, 463

\reference {} Norris, J.E., Christlieb, N., Korn, A.J., Eriksson, K.,
Bessell, M.S., Beers, T.C., Wisotzki, L., Reimers, D. 2007, ApJ, 670,
774

\reference {} Norris, J.E. \& Da Costa, G.S. 1995, ApJ, 447, 680

\reference {} Norris, J.E., Freeman, K.C., \& Mighell, K.J. 1996, ApJ,
462, 241

\reference {} Norris, J.E., Gilmore, G., Wyse, R.F.G.,
Wilkinson, M.I., Belokurov, V., Evans, N.W., Zucker, D.B. 2008, ApJ,
689, L113

\reference {} Norris, J.E., Ryan, S.G., \& Beers, T.C. 1997, ApJ, 489,
L169

\reference {} Norris, J.E., Yong, D., Gilmore, G., \& Wyse,
R.F.G. 2010, ApJ, 711, 350

\reference {} Pancino, E., Pasquini, L., Hill, V., Ferraro, F.R., \&
Bellazzini, M. 2002, ApJ, 568, L101

\reference {} Persson, S.E., Frogel, J.A., Cohen, J.G., Aaronson, M.,
\& Matthews, K. 1980, ApJ, 235, 452

\reference {} Plez, B. \& Cohen, J.G. 2005, A\&A, 434, 1117

\reference {} Renzini, A. \& Fusi Pecci, F. 1988, ARA\&A, 26, 199

\reference {} Robin, A.C., Reyl\'{e}, C., Derri\`{e}re, S., \& Picaud,
S. 2003, A\&A, 409, 523 Erratum: 2004, A\&A, 416, 157

\reference {} Ryan, S.G., Norris, J.E., \& Beers, T.C. 1996, ApJ, 471,
254

\reference {} Sneden. C. 1973, ApJ, 184, 839


\reference {} Spite, M. et al.\ 2005, A\&A, 430, 655

\reference {} Stanford, L.M., Da Costa, G.S., Norris, J.E., \& Cannon,
R.D. 2007, ApJ, 667, 911

\reference {} Tolstoy, E., Hill, V., \& Tosi, M. 2009, ARA\&A, 47, 371

\reference {} Tolstoy, E. et al.\ 2004, ApJ, 617, L119

\reference {} Trefzger, C.F., Carbon, D.F., Langer, G.E., Suntzeff,
N.B., \& Kraft, R.P. 1983, ApJ, 266, 144

\reference {} Vivas, A.K., Zinn, R., \& Gallart, C. 2005, AJ, 129, 189


\reference {} Winnick, R.A. 2003, PhD thesis, Yale University


\reference {} Zhao, C. \& Newberg, H.J. 2006, astro-ph/0612034

\end{references}
\end{document}